\documentclass[aip,preprint, amsmath,amssymb,MnSymbol,mathtools]{revtex4-1}

\usepackage{textcmds}
\usepackage{xr}
\usepackage{float}
\usepackage{graphicx}
\usepackage{dcolumn}
\usepackage{bm}
\usepackage[mathlines]{lineno}

\usepackage[utf8]{inputenc}
\usepackage[T1]{fontenc}
\usepackage{mathptmx}
\usepackage{etoolbox}
\usepackage[T1]{fontenc} 
\ProvideTextCommand{\DJ}{OT1}{\raisebox{0.25ex}{-}\kern-0.45em D}
\usepackage{booktabs}
\draft 

\begin{document}


\title{Stress-controlled medium-amplitude oscillatory shear (MAOStress) of PVA-Borax} 



\author{Nabil Ramlawi}
\affiliation{Department of Mechanical Science and Engineering, University of Illinois Urbana-Champaign, Urbana, Illinois 61801, USA}

\author{Mohammad Tanver Hossain}
\affiliation{Department of Mechanical Science and Engineering, University of Illinois Urbana-Champaign, Urbana, Illinois 61801, USA}
\author{Abhishek Shetty}
\affiliation{Rheology Division, Anton Paar USA, 10215 Timber Ridge Dr, Ashland, VA, 23005, USA}
\author{Randy H. Ewoldt}
\affiliation{Department of Mechanical Science and Engineering, University of Illinois Urbana-Champaign, Urbana, Illinois 61801, USA}
\affiliation{Materials Research Laboratory, University of Illinois at Urbana-Champaign, Urbana, Illinois 61801, USA}


\date{\today}

\begin{abstract}
We report the first-ever complete measurement of MAOStress material functions, which reveal that stress can be more fundamental than strain or strain rate for understanding linearity limits as a function of Deborah number. The material used is a canonical viscoelastic liquid with a single dominant relaxation time: Polyvinyl alcohol (PVA) polymer solution crosslinked with tetrahydroborate (Borax) solution. We outline experimental limit lines and their dependence on geometry and test conditions. These MAOStress measurements enable us to observe the frequency dependence of the weakly nonlinear deviation as a function of stress amplitude. The observed features of MAOStress material functions are distinctly simpler than MAOStrain, where the frequency dependence is much more dramatic. The strain-stiffening transient network model (SSTNM) was used to derive a model-informed normalization of the nonlinear material functions that accounts for their scaling with the linear material properties. Moreover, we compare the frequency-dependence of the critical stress, strain, and strain-rate for the linearity limit, which are rigorously computed from the MAOStress and MAOStrain material functions. While critical strain and strain-rate change by orders of magnitude throughout the Deborah number range, the critical stress changes by a factor of about two, showing that stress is a more fundamental measure of nonlinearity strength. This work extends the experimental accessibility of the weakly nonlinear regime to stress-controlled instruments and deformations, which reveal material physics beyond linear viscoelasticity but at conditions that are accessible to theory and detailed simulation.  
\end{abstract}


\maketitle 


\section{Introduction}
Measuring the nonlinear viscoelastic response of complex fluids provides valuable insights into their constitutive behavior and underlying physics. One technique for studying the nonlinear response is large-amplitude oscillatory shear (LAOS)\cite{Hyun2011ALAOS,Giacomin1993Large-amplitudeShear,Wilhelm2002FourierTransformRheology,Ewoldt2008NewShear}, which has been used to study a wide range of complex materials\cite{Komatsu1973NonlinearEmulsions,Neidhofer2004DistinguishingRheology,KateGurnon2012LargeMicelles,Dealy2013MeltIndustry,Randall2014Linear-nonlinearPolymers,Joyner2021NonlinearQuality}. However, this technique can be hindered by noisy data, sample issues, and difficulty in relating the results to models. Weakly nonlinear methods, such as medium-amplitude oscillatory shear (MAOS)\cite{Hyun2009EstablishingSystems,Ewoldt2013Low-dimensionalViscoelasticity, KateGurnon2012LargeMicelles,Kumar2016IntrinsicSuspensions,Singh2017Frequency-sweepMAOS,Bharadwaj2017AShear,Carey-DeLaTorre2018First-harmonicMAOS,Natalia2020QuestioningExpansions} and medium amplitude superposition (MAPS)\cite{Lennon2020MediumExamples,Shanbhag2022KramersKronigImplications}, offer a sweet spot between small-amplitude oscillatory shear (SAOS) and LAOS probing a regime with rich detail that is theoretically tractable.

\begin{figure}[h!]
 \includegraphics{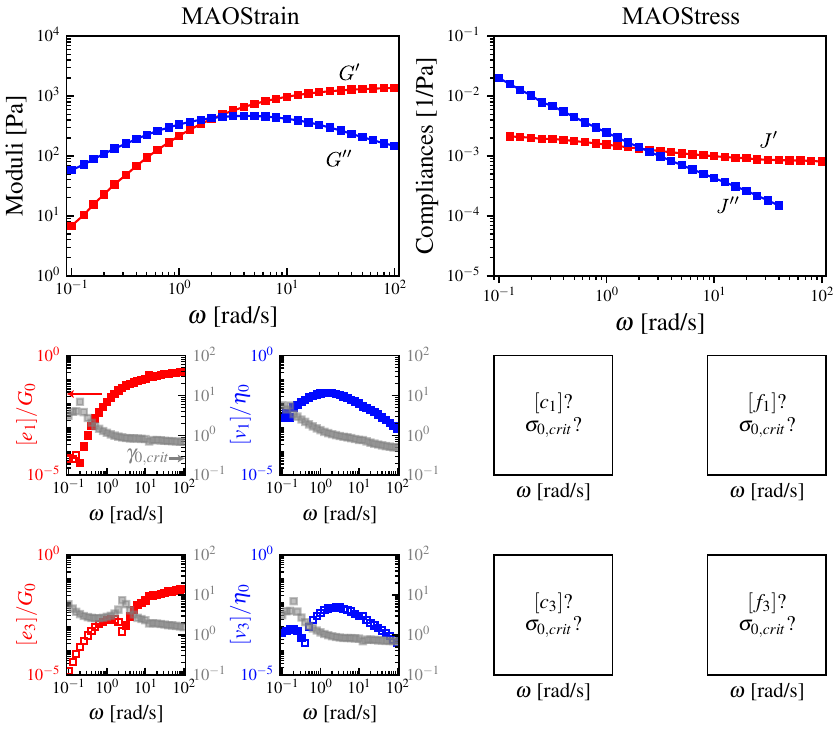}%
 \caption{MAOStrain vs MAOStress. Linear moduli and compliances are interchangeable but exhibit different features. Nonlinear MAOStrain material functions ($[e_1]$, $[v_1]$, $[e_3]$, and $[v_3]$) were reported previously\cite{Bharadwaj2017AShear}, but we repeat experiments here for consistency (Appendix C). The critical strain for nonlinearity can be calculated from the MAOStrain material functions(Eq. 5 in Ewoldt and Bharadwaj\cite{Ewoldt2013Low-dimensionalViscoelasticity}, with $\epsilon=10\%$). This work aims to measure the dependence of MAOStress material functions on frequency for the first time, revealing how the critical stress to nonlinearity depends on frequency. The data used here is for the 2.75$\%$PVA/1.25$\%$Borax sample investigated in this work.}
 \label{fig:3intro}
 \end{figure}
 
MAOS probes the gradual deviation from linearity in an oscillatory test. For example, applying a sinusoidal strain input represented as $\gamma(t)=\gamma_0\sin \omega t$ in strain-controlled MAOS (MAOStrain) results in a time-periodic stress response that can be written as the expansion
\begin{align}
\sigma (t)=\gamma_0\left( \right.& \left.G^{\prime}\sin\omega t+G^{\prime\prime}\cos \omega t \right) \nonumber\\
+\gamma_0^3& \left( [e_1]\sin \omega t+[v_1]\omega \cos \omega t\right.  \\
-&\left. [e_3]\sin 3 \omega t+[v_3] \omega \cos 3\omega t\right) + \mathcal{O}(\gamma_0^5),\nonumber
\end{align}
where $[e_1]$, $[v_1]$, $[e_3]$, and $[v_3]$ are the frequency-dependent intrinsic Chebyshev nonlinearities defined by Ewoldt et al. \cite{Ewoldt2013Low-dimensionalViscoelasticity}. Measuring these material functions provides more information about the material, quantifying changes in the nonlinear material response as a function of frequency (e.g., Fig.~\ref{fig:3intro}). This regime is also accessible to analytical model solutions\cite{Bharadwaj2015ConstitutiveShear,Saengow2017ExactStress,Martinetti2019Time-strainShearb,Bharadwaj2017AShear,Song2020EvaluatingSolutions,Ramlawi2020TheModel}, which allows for material-property inference using information from both the linear and nonlinear material response\cite{Singh2022OnCertainty}. MAOStrain, for instance, has been used to solve a 70-year-long debate in the literature about the origin of strain-stiffening in transient polymer networks\cite{Martinetti2018InferringNetworkb}. Moreover, the MAOS material functions can be used to calculate the linearity limits of viscoelastic materials as a function of frequency by quantifying the absolute strength of the nonlinear deviation compared to the linear term. The linearity limits can be used to describe the regime boundary on a Pipkin map\cite{Pipkin1986LecturesTheory,Dealy1999MeltProcessing,Ewoldt2013Low-dimensionalViscoelasticity,Giacomin2011Large-amplitudeModel} that separates the linear and nonlinear regimes~\cite{Davis1978NonlinearPMMA,Jongschaap1978OnDisks,Pearson1982BehaviorFields}.

The MAOS approach to compute linearity limits in an oscillatory test is more rigorous than using the total deviation in the first harmonics. Since the nonlinear transition is gradual, linearity limits are typically defined as the forcing amplitude (stress, strain, or strain rate) at which the deviation from the linear plateau of a nonlinear material function exceeds a subjective threshold $\epsilon$ (e.g., $\frac{|G_1^{\prime}(\gamma_0)-G^{\prime}|}{G^{\prime}}>\epsilon$)\cite{Pearson1982BehaviorFields}. The MAOS material functions quantify this gradual transition\cite{Ewoldt2013Low-dimensionalViscoelasticity} (e.g., $G_1^{\prime}=G^{\prime}+[e_1]\gamma_0^2+...$ for small-enough amplitudes), and can be used to define a critical forcing amplitude for an arbitrary threshold (e.g., $\frac{\gamma_{0,crit}(\omega)}{\sqrt{\epsilon}}=\sqrt{\frac{G^{\prime}(\omega)}{[e_1](\omega)}}$ and $\dot{\gamma}_{0,crit}=\omega\gamma_{0,crit}$), as shown in Fig.~\ref{fig:3intro}. The benefit of using the MAOS definition instead of the total deviation is that the total deviation may not be measurable for small threshold values (the corresponding forcing amplitude is too small), unlike the MAOS deviation that is quantifiable at any forcing amplitude if the material function is measured. 

Nonlinear oscillatory measurements can be conducted in both strain-control and stress-control modes. While most of the LAOS measurements in the literature are strain-controlled (LAOStrain), many works used LAOStress (stress-controlled LAOS) to explore the nonlinear response of materials\cite{Dimitriou2013DescribingLAOStress, Pineiro-Lago2023LargePDO}. Stress-controlled results can be connected directly to many practical applications, such as gravity and pressure-driven flows. For certain materials, such as yield-stress fluids, stress-controlled oscillations allow for direct probing of the yielding transition and can add to the understanding derived from LAOStrain measurements. Moreover, Lauger and Stettin showed that LAOStrain and LAOStress probe a different material behavior, where significant differences in the Lissajous stress-strain curves were observed for Xantham gum\cite{Lauger2010DifferencesFluids}. Hence, a better understanding of stress-controlled nonlinear measurements will increase the accessibility of MAOS measurements. 

MAOStress has been theoretically anticipated since at least 2013 but has yet to be experimentally measured. A theoretical framework for stress-controlled MAOS (MAOStress) was developed by Ewoldt and Bharadwaj in 2013. For a sinusoidal stress input represented as $\sigma (t)=\sigma_0\cos \omega t$, the time-periodic strain response in the weakly nonlinear regime can be expanded as 
\begin{align}
\gamma(t)=\bar{\gamma}(\omega,\sigma_0)+\sigma_0  &\left(J^{\prime}\cos\omega t+J^{\prime\prime}\sin \omega t\right)\nonumber\\
+\sigma_0^3&\left( [c_1]\cos \omega t+\frac{[f_1]}{\omega} \sin \omega t\right.\\
+&\left.[c_3]\cos 3 \omega t+\frac{[f_3]}{3\omega}\sin 3\omega t\right) + \mathcal{O}(\sigma_0^5),\nonumber
\end{align}
where $[c_1]$, $[f_1]$, $[c_3]$, and $[f_3]$ are the frequency-dependent intrinsic Chebyshev nonlinear compliances and fluidities and $\bar{\gamma}(\omega,\sigma_0)$ is the baseline strain, as defined by Ewoldt et al. \cite{Ewoldt2013Low-dimensionalViscoelasticity}. Similar to MAOStrain, a critical stress amplitude for nonlinearity $\sigma_{0,crit}$ can also be defined based on the measured MAOS material functions. While many MAOStrain measurements have been reported, no frequency-dependent MAOStress material functions have been previously reported.  

In this work, we make the first-ever MAOStress measurement to test the weakly nonlinear response of PVA-Borax, including $\gamma_{0,crit}$, $\dot{\gamma}_{0,crit}$, and $\sigma_{0,crit}$. PVA-Borax has been well-studied using MAOStrain\cite{Bharadwaj2017AShear,Martinetti2018InferringNetworkb}, making it a model material to compare the two protocols. In the linear regime, stress- and strain-control protocols are interchangeable, but the stress-controlled perspective can still show a much simpler response (Fig.~\ref{fig:3intro}). Therefore, we want to examine how the material response signatures differ between these two methods. With the four MAOStrian and MAOstress material functions, one can identify the frequency-dependent limits of the linear viscoelastic approximation (i.e., the strength of nonlinearity, which can be reinterpreted as a critical forcing strength where nonlinearity is \q{significant} in terms of critical strain amplitude, or strain rate amplitude, or stress amplitude). We compare the frequency dependence of critical stress, strain, and strain rate for the emergence of nonlinearity. We show that stress is a more universal measure of nonlinearity strength, where the critical stress for nonlinearity varies minimally with frequency compared to the critical strain and strain rate.
 
This work is organized as follows: First, we introduce the material preparation protocol, the theoretical framework for MAOStress, and the experimental procedure that was followed. Section 3 presents the experimental measurement of the MAOStress material functions, where the frequency dependence of the weakly nonlinear response is thoroughly analyzed, in addition to analysis of the scaling of the nonlinear material functions with linear properties. Section 4 discusses the implications of the results on the frequency dependence of the critical stress, showing how Pipkin map limit lines can be affected by the choice of strain, strain rate, or stress as the measure of the forcing amplitude.

\label{Sec: Intro}

\section{Materials and Methods}

 \begin{figure}[h!]
 \includegraphics[width=\textwidth]{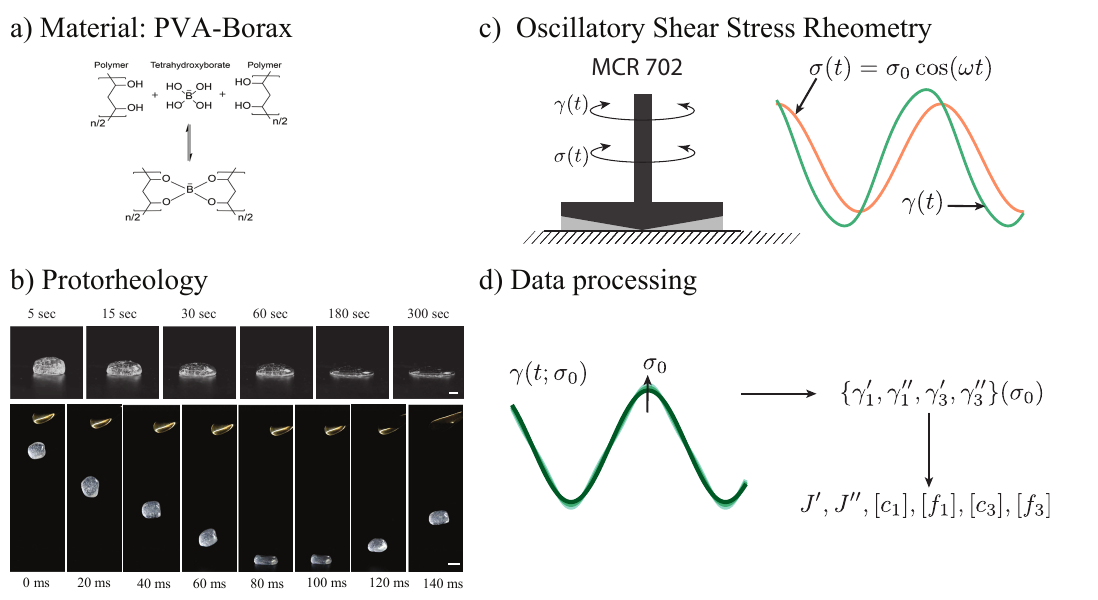}%
 \caption{Materials and methods. (a) PVA-Borax is a transient polymer network with relaxation time $\tau \approx 0.4$s, allowing it to (b) bounce upon impact (short times) while creeping under its weight at long times as assessed by protorheology\cite{Hossain2024Protorheology}. (c) Cone-plate geometry was used on a stress-controlled rheometer to impose a stress co-sinusoidal input signal. The directly measurable quantities are torque $M(t)$ and angular displacement $\theta(t)$, which are proportional to stress ($\sigma(t)=\frac{2}{\pi R^3}M(t)$) and strain ($\gamma(t)=\frac{1}{\tan \alpha}\theta(t)$) respectively, where $R$ and $\alpha$ are the radius and angle of the cone, respectively. (d) The measured strain waveform will deviate from the ideal sinusoidal shape at increasing stress amplitude $\sigma_0$. Discrete Fourier transform are used to identify the strain harmonics $\{\gamma_1^{\prime},\gamma_1^{\prime\prime},\gamma_3^{\prime},\gamma_3^{\prime\prime}\}$  (Eq.~\ref{eq:3 Fourier expansion}) at every amplitude, which are fit to Eq.~\ref{eq:3 harmnonics expansion} to find the MAOStress material functions at each frequency.  All scale bars are 1 cm. \label{fig:3Materials and Methods}}%
 \end{figure}
 
\subsection{Mathematical Framework}
In simple shear deformation, the time-periodic strain response to an oscillatory shear stress, represented as $\sigma (t)=\sigma_0\cos \omega t$ for convenience \cite{Ewoldt2013DefiningShear}, can be written as a Fourier expansion in terms of the odd harmonics\cite{Ewoldt2013DefiningShear,Hyun2011ALAOS}, in addition to a baseline strain $\bar{\gamma}$
\begin{equation}
\label{eq:3 Fourier expansion}
    \gamma(t;\sigma_0,\omega)=\bar{\gamma}(\sigma_0,\omega)+\sum_{n,\mathrm{odd}} \gamma_n^{\prime}(\sigma_0,\omega)\cos n\omega t+\gamma_n^{\prime\prime}(\sigma_0,\omega)\sin n\omega t,
\end{equation}
where $\gamma_n^{\prime}$ and $\gamma_n^{\prime\prime}$ are the $\mathrm{n^{th}}$ elastic and viscous strain harmonics, respectively. For sufficiently small amplitudes, the strain behaves linearly, where only the first harmonics, $\gamma_1^{\prime} $ and $\gamma_1^{\prime\prime}$, are required to describe the response, and these harmonics scale linearly with stress amplitude. In this limit of $\sigma_0 \to 0$, referred to as small-amplitude oscillatory shear (SAOS), the stress-independent linear elastic compliance $J^{\prime}(\omega)=\gamma_1^{\prime}/\sigma_0 $ and linear viscous compliance $J^{\prime\prime}(\omega)=\gamma_1^{\prime\prime}/\sigma_0 $ (or fluidity $\phi^{\prime}=J^{\prime\prime}\omega$) capture the frequency-dependence of the viscoelastic behavior. However, as the amplitude increases, the linearity with stress amplitude is broken by the emergence of higher harmonics and changes to the now amplitude-dependent first harmonics. One common assumption is that these harmonics can be expanded as a function of the stress amplitude $\sigma_0$ using an integer power series, which applies to many materials. Nevertheless, some materials have shown non-integer power expansions\cite{Natalia2020QuestioningExpansions, Natalia2022}, which can be predicted from physical models such as Hertzian particle contact\cite{Natalia2022}. However, since previous experimental observations of PVA-Borax showed that an integer power series expansion works for that material, we expand the strain harmonics as integer powers. 
In the first deviation from linearity, only the first and third harmonics are present\cite{Ewoldt2013Low-dimensionalViscoelasticity}, and the first and third strain harmonics up to the first deviation are expanded as
\begin{subequations}
\label{eq:3 harmnonics expansion}
\begin{alignat}{2}
\gamma_1^{\prime}(\sigma_0,\omega) &=J^{\prime}\sigma_0 + &[c_1]\sigma_0^3+ \mathcal{O}(\sigma_0^5)\\
\gamma_1^{\prime\prime}(\sigma_0,\omega) &=J^{\prime\prime}\sigma_0 + &\frac{[f_1]}{\omega}\sigma_0^3+ \mathcal{O}(\sigma_0^5)\\
\gamma_3^{\prime}(\sigma_0,\omega) &= &[c_3]\sigma_0^3+ \mathcal{O}(\sigma_0^5)\\
\gamma_3^{\prime\prime}(\sigma_0,\omega) &= &\frac{[f_3]}{3\omega}\sigma_0^3+ \mathcal{O}(\sigma_0^5).
\end{alignat}
\end{subequations}
The first-harmonic intrinsic nonlinearities, $[c_1]$ and $[f_1]$, quantify the degree to which changes in the first harmonics occur as the stress amplitude increases, causing intracycle changes in the material response. The sign of these material functions determines whether softening, stiffening, thinning, or thickening occurs as amplitude increases. The third harmonic intrinsic nonlinearities $[c_3]$ and $[f_3]$ quantify the leading-order growth of the third harmonics of strain, causing increased distortion in the Lissajous curves.

\subsection{Experimental Protocol}
The stress-controlled oscillatory amplitude sweeps were conducted on the MCR 702 rotational rheometer in combined motor transducer mode. Care was taken to avoid experimental limitations\cite{Ewoldt2015ExperimentalData}. We used a cone and plate geometry ($d = 25 \  \mathrm{mm}$, $\alpha=4^{\circ}$) to ensure homogeneous stress and strain distribution within the sample. The temperature was controlled using a bottom Peltier plate at $25^{\circ}\mathrm{C}$. After loading, the sample was coated with a thin layer of silicone oil (100~ cst, Sigma Aldrich) at the edge of the gap to prevent water evaporation from the sample. 

The material functions were extracted at every frequency by fitting the stress amplitude sweeps to Eqs.~\ref{eq:3 harmnonics expansion}(a)-(d). The maximum stress amplitude tested has to be high enough to resolve the MAOS regime experimentally. However, it is preferred to avoid high stresses beyond the MAOS regime, which may cause irreversible changes to the material. This allows us to use the same sample to test multiple frequencies, simplifying the experimental protocol. Hence, several pretests were run to find the maximum stress amplitude we can reach at every frequency without causing irreversible change by comparing the linear viscoelastic frequency sweep before and after the sweep. This maximum amplitude and the minimum amplitude required to resolve the MAOS regime experimentally vary as a function of frequency. Therefore, a different stress range is set at each frequency. Moreover, the time required to complete the sweeps at the low frequency can exceed 24 hours, which puts the sample at risk of evaporation despite usingcone oil. For this reason, multiple loadings were used to test the entire frequency range.

The full frequency range ($\omega=[0.1-100] \mathrm{rad/s}$)  was tested in three different loadings. We used $\sigma_0=[1-300]$Pa for $\omega=[0.1-1] \mathrm{rad/s}$, $\sigma_0=[1-500]$Pa for $\omega=[1.0-10.0] \mathrm{rad/s}$, and $\sigma_0=[1-1000]$Pa for $\omega=[10.0-100.0] \mathrm{rad/s}$. Moreover, we used 50 points per decade of stress amplitude $\sigma_0$ to get good resolution and to ensure that the steps in stress amplitude are small, such that the equilibration time from one amplitude to the other is low. 
 \begin{figure}[h!]
 \includegraphics[width=\textwidth]{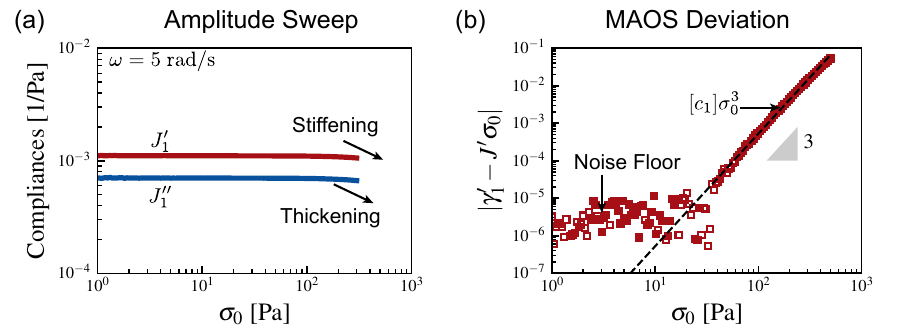}%
 \caption{(a) Amplitude sweep of first-harmonic compliances $J_1^{\prime}$ and $J_1^{\prime\prime}$ showing the stiffening and thickening occurring in PVA-Borax at $\omega=5$ rad/s in the range of stress amplitude tested. (b) The weakly nonlinear deviation of $J_1^{\prime}$ can be isolated and fit after subtracting the linear term $J^{\prime}\sigma_0$. The angular displacement noise floor corresponding to strain differences of around $~10^{-5}-10^{-6}$ is due to instrument limitations and ambient noise.}
 \label{fig:3Sweep Illustration}
 \end{figure}

A custom Python code is used to process the stress and strain waveforms at every amplitude and frequency using a discrete Fourier transform to extract the strain harmonics $\{\gamma_1^{\prime}, \;\gamma_1^{\prime\prime},\; \gamma_3^{\prime}, \;\gamma_3^{\prime\prime}\}(\sigma_0,\omega)$. The amplitude sweeps of the first-harmonic compliances ($J_1^{\prime}$ and $J_1^{\prime\prime}$) can be used to identify the overall response of the material to the applied stress. An amplitude sweep for PVA-Borax is shown in Fig.~\ref{fig:3Sweep Illustration} at $\omega=5$ rad/s. The decrease in compliances at increasing amplitudes highlights these networks' stiffening and thickening nature. While the linear compliances can be extracted from the first-harmonic compliances sweep, the intrinsic nonlinear coefficients are computed by isolating the nonlinear deviations as shown in Fig.~\ref{fig:3Sweep Illustration}b. The linear term can be subtracted from the elastic first-harmonic in Eq.~\ref{eq:3 harmnonics expansion}a to isolate the weakly nonlinear term. At every frequency, the four weakly nonlinear deviations ($\{\gamma_1^{\prime}, \;\gamma_1^{\prime\prime},\; \gamma_3^{\prime}, \;\gamma_3^{\prime\prime}\}(\sigma_0,\omega)$) are fit to Eq.~\ref{eq:3 harmnonics expansion} to extract the MAOStress material functions $J^{\prime}$, $J^{\prime\prime}$, $[c_1]$, $[c_3]$, $[f_1]$, and $[f_3]$. The range of the fit was restricted to the experimental MAOS regime, which is lower bounded by the noise floor and upper bounded by the higher order terms (>$\sigma_0^3$). Quantifying weak deviations is non-trivial, and we fully document the fits and the corresponding ranges in the Supplementary Information (SI). 

\begin{figure}[h!]
\centering
 \includegraphics{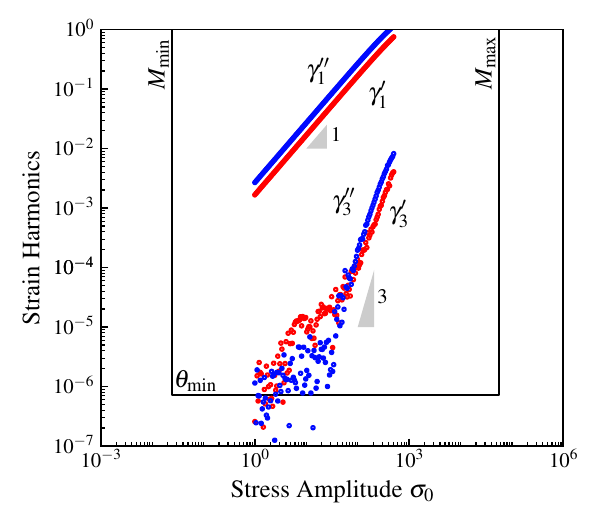}%
 \caption{Experimental limit lines for stress amplitude sweeps. Limit lines for low-torque (Eq.~\ref{eq:3Mmin}), high-torque (Eq.~\ref{eq:3Mmax}), and minimum angular displacement (Eq.\ref{eq:3thetaMin}) plotted for 25mm cone-plate using the instrument specs in Table \ref{tab:MCRspecs}. PVA-Borax data collected at 1 rad/s shows that the true limit line for angular displacement is about one order of magnitude higher than the theoretical limit line. \label{fig:3Experimental limits}}%
 \end{figure}

Experimental limits\cite{Ewoldt2015ExperimentalData} for the measured harmonics and compliances are useful to determine the accessible experimental window. From the primary observables of torque and rotational displacement, the window is set by the low torque limit $M_{\mathrm{min}}$
\begin{equation}
    {
    \label{eq:3Mmin}
 \sigma_{\mathrm{min}}=F_\sigma M_{\mathrm{min}},
}
\end{equation}
the high torque limit $M_{\mathrm{max}}$
\begin{equation}
{
\label{eq:3Mmax}
     \sigma_{\mathrm{max}}=F_\sigma M_{\mathrm{max}},
}
\end{equation}
and the minimum displacement limit $\theta_{\mathrm{min}}$ 
\begin{equation}
{\label{eq:3thetaMin}
\gamma_{\mathrm{min}}=F_{\gamma}\theta_{\mathrm{min}} }
\end{equation}
of the instrument. For a cone and plate geometry, $F_{\sigma}=\frac{2}{\pi R^3}$ and $F_{\gamma}=\frac{1}{\tan \alpha}$, where $R$ is the radius of the geometry and $\alpha$ is the degree of the cone. The minimum measurable compliance can be inferred from the minimum strain as $J_{\mathrm{min}}=\gamma_{\mathrm{min}}/\sigma_0$. Figure~\ref{fig:3Experimental limits} demonstrates how the experimental limits can be used to identify the theoretically accessible MAOStress experimental window of the instrument, e.g., based on the manufacturer specifications shown in Table \ref{tab:MCRspecs}. The predicted low-displacement limit is only an order of magnitude higher than the measured limit of $\gamma_{\mathrm{min}}=10^{-5}$. Moreover, the weakly nonlinear deviation characterized by the slope of 3 for the third harmonics in Fig.~\ref{fig:3Experimental limits} can be significantly below the high-torque limit of the instrument. Despite that, the predicted low-torque limit is $\sigma_{\mathrm{min}}=10^{-2}$ Pa, and the raw waveforms were found to be noisy below $\sigma_0=10$~Pa. Therefore, the low-stress limit from Eq.~\ref{eq:3Mmin} may not directly translate to the stress amplitude that can be measured. One possible explanation is that the torque is controlled in steps much smaller than the stress amplitude during an oscillatory test. For example, for the 512 points per cycle used, the minimum stress change between successive points can be found at the peak of the oscillatory wave. It can be estimated as $\frac{\Delta \sigma_{min}}{\sigma_0} =|\cos(0)-\cos(\Delta t\omega)|$, where $\Delta t\omega=2\pi/512$ for a fixed time interval between points. Therefore, the minimum stress amplitude that can be controlled should be $\sigma_{0,\mathrm{min}}=\frac{\Delta \sigma_{min}}{1-\cos(2\pi/512)}\approx \frac{F_{\sigma}M_{\max}}{7*10^{-5}}=4$ Pa for the geometry used in this work. This value provides a good starting point for experimental design, where the difference between the predicted (4 Pa) and actual (around 10 Pa from Fig.~\ref{fig:3Experimental limits}) low-stress limit is around a factor of 2 and can be due to ambient noise.  

\begin{table}
\centering
\caption{\label{tab:MCRspecs}MCR 702 manufacturer instrument specification.}
\begin{tabular}{ >{\raggedright\arraybackslash}p{0.1\linewidth}  >{\raggedright\arraybackslash}p{0.1\linewidth}>{\raggedright\arraybackslash}p{0.1\linewidth} }
 $M_{\mathrm{min}}$ & $M_{\mathrm{max}}$&
$\theta_{\mathrm{min}}$\\
\hline
0.5nN.m & 230mN.m & 50nrad \\
\end{tabular}

\end{table}
\subsection{Material Preparation}
The PVA-Borax sample in this work contains 2.75 wt\% Polyvinyl alcohol (PVA) aqueous solution physically crosslinked with 1.25 wt\% Sodium tetraborate (Borax). The concentration was selected following the work of Bharadwaj et al.\cite{Bharadwaj2017AShear}. The sample was prepared as follows: 4wt\% PVA (Sigma-Aldrich, 99\% hydrolyzed, Mw = 85,000-110,000) was dissolved in DI water at $80^\circ \mathrm{C}$ overnight until a homogeneous transparent solution was obtained. Similarly, 4wt\% Borax (Sigma-Aldrich) solution was prepared by dissolving Borax in DI water. After the stock solutions cooled down to room temperature, the two solutions were mixed (27.5 PVA:12.5 Borax by solution mass) to prepare 2.75wt\% PVA- 1.25wt\% Borax solution. Air bubbles trapped during the mixing were removed by centrifuging.

Protorheology\cite{Hossain2024Protorheology}, shown in Fig.~\ref{fig:3Materials and Methods}(b), reveals that this PVA-Borax composition behaves as a viscoelastic liquid with a finite flow viscosity $\eta$, observable elasticity, and relaxation time $\tau$. Viscous flow is revealed by the time-lapse images, with observable flow on the order of tens of seconds and proceeding for 300~s in the images. A quantitative estimate of viscosity can be made by considering the radial evolution between $t=0$~s and $t=180$~s, assuming a viscous gravity current with steady viscous flow and negligible surface tension, as $\eta=0.0195\rho g \Delta t\frac{h_2^3(3R_2^2+h_2^2)^3}{R_2^8-R_1^8}$ where $\rho$ is density, $\Delta t$ is elapsed time, $h_2$ is final height and $R_1$, $R_2$ are the initial and final radius, respectively. The $t=300$~s frame was not considered because surface tension effects may compete with the viscous effects at that low height, and only viscosity and gravity were considered in the protorheology analysis. From Fig.~\ref{fig:3Materials and Methods}(b), we have $R_1$ = 1.6~cm, $R_2$~=~2.5~cm and $h_2$~= 1.1~cm and using $\rho = 10^3 
 \ \mathrm{kg/m^3}$ and g =~9.8 N/kg, we obtained $\eta \approx 2000 \;\mathrm{Pa.s}$ at a characteristic stress of $\sigma_{grav} \approx \;220 \ \mathrm{Pa}$ and  observation time of $t_{obs} = 180$~s. This stress, as will be shown below, is within the weakly-nonlinear regime, which suffices in providing an approximation of the zero-frequency fluidity $\phi_0\approx\frac{1}{\eta}=5\times 10^{-4}$~$(\mathrm{Pa.s})^{-1}$, which is within an order of magnitude of the measured $\phi_0$.
 The elastic behavior of PVA-Borax is indicated by the bounce test in Fig.~\ref{fig:3Materials and Methods}(b) which shows elasticity at short times on the order of tens of milliseconds and a small loss factor $\tan\delta=G^{\prime\prime}/G^{\prime}<1$ at these bouncing conditions, where $\tan\delta$ can be quantitatively estimated from $\tan \delta=\frac{1}{\pi}ln(\frac{1}{f})$, where $f=h_{reb}/h_{drop}$ is the recovered height fraction. From Fig.~\ref{fig:3Materials and Methods}(b) we have $f = 0.38$ which gives $\tan\delta \approx0.3$ (corresponding to $\omega=\pi/\Delta t \approx157$ rad/s where $\Delta t \approx 20$ ms) at a characteristic stress of $\sigma\approx210$~Pa. The relaxation time $\tau$ is quantitatively bounded from these observations, being between the bouncing and time-lapse time scales, i.e., greater than 20~ms and less than 15~s. The protorheology observations are consistent with the linear viscoelastic frequency sweeps in Fig.~\ref{fig:3intro} and provide compelling visual interpretations of key features therein. The images in Fig.~\ref{fig:3Materials and Methods}(b) do not indicate the nature of the nonlinearity, which will be revealed by precision rheometry. 

\section{Results}
\begin{figure}[H]
{ \includegraphics[width=\textwidth]{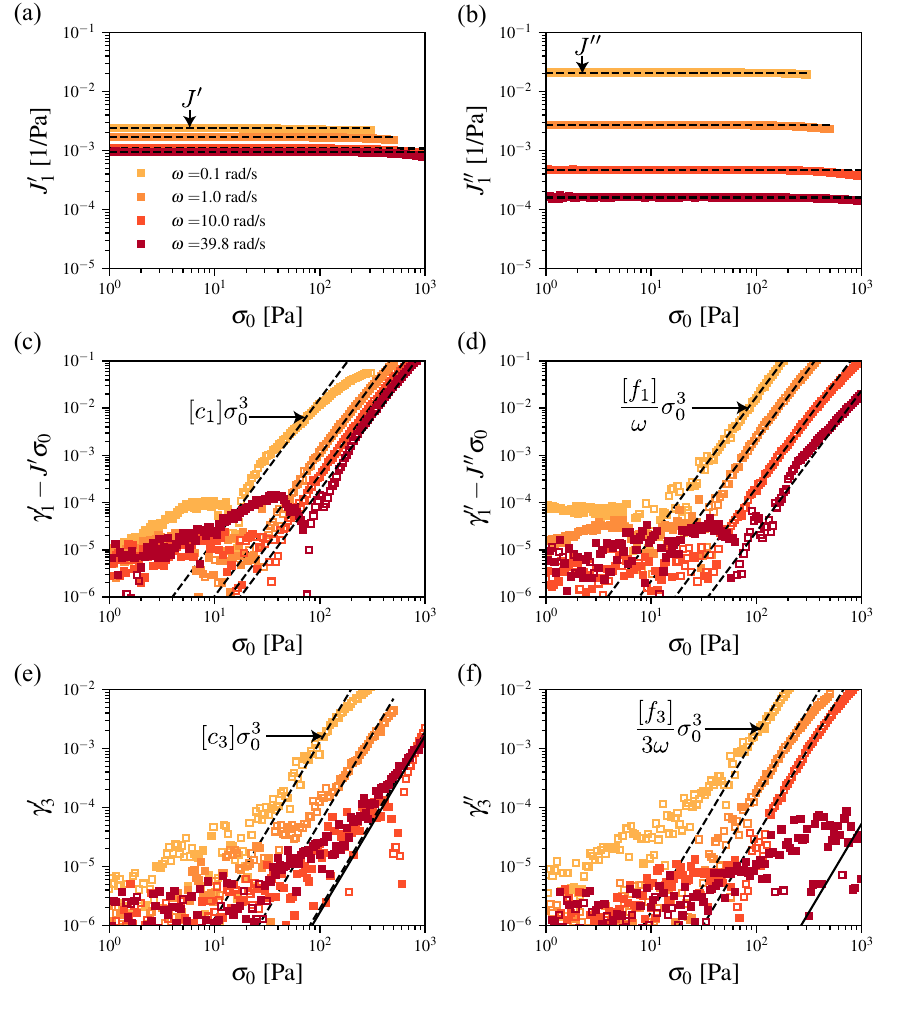}%
 \caption{Stress amplitude sweeps of (a) elatsic $J_1^{\prime}$ and (b) viscous $J_1^{\prime\prime}$ first-harmonic compliances at $\omega={0.1, 1, 10, 39.9}$rad/s.  Dashed lines show $J^{\prime}$ and $J^{\prime\prime}$ values obtained from the fits. The corresponding growth of the elastic (c) and viscous (d) first-harmonic deviations are with a power-law of 3 (Eq.\ref{eq:3 harmnonics expansion}). Similarly, elastic (e) and viscous (f) third harmonics' leading order term fit with a power law of 3. Closed symbols represent positive values, and open symbols represent negative values. Similar plots for all frequencies are shown in the supplementary information (Figs. S2-S32)}%
  \label{fig:3AmpSweep}}
 \end{figure}

Stress amplitude sweeps of PVA-Borax were conducted in the frequency range of [0.1-100 rad/s], and the resulting first and third strain harmonics are shown in Fig.~\ref{fig:3AmpSweep} for frequencies of $\omega=[0.1,1,10,39.8]$ rad/s. The first-harmonic compliances exhibited a typical linear regime plateau, followed by a slight decrease due to the stress-stiffening nature of the material. The slight decrease in compliance led to orders of magnitude growth in the weakly nonlinear deviations of the first harmonics. These deviations were computed by subtracting the linear contributions, calculated as $J^{\prime}\sigma_0$ and $J^{\prime\prime}\sigma_0$ for the elastic and viscous first harmonics. The fit lines in Fig.~\ref{fig:3AmpSweep}(c)-(f) demonstrate that the first and third-harmonic deviations grew closely to $\sigma_0^3$, consistent with the theory and measurements of the weakly nonlinear deviations in MAOStrain. The material functions associated with the fit lines were extracted as described in the methods section, and the fits for all the frequencies are shown in the SI (Figs.S2-S32). The observable deviations grow after a certain amplitude where the signal exceeds the noise floor. This noise was above the value predicted from the instrument limits using eq.~\ref{eq:3thetaMin}. Furthermore, the noise floor increased with the experimental frequency. These observations can be explained by external factors, such as ambient noise, instrument state, and sample loading, that are difficult to predict a priori and are not accounted for by instrument calibration. Nonetheless, the minimum observable strain amplitude achieved was remarkably low ($<10^{-5}$) and was enough to extract the MAOStress material functions.

\begin{figure}[h!]
 \includegraphics{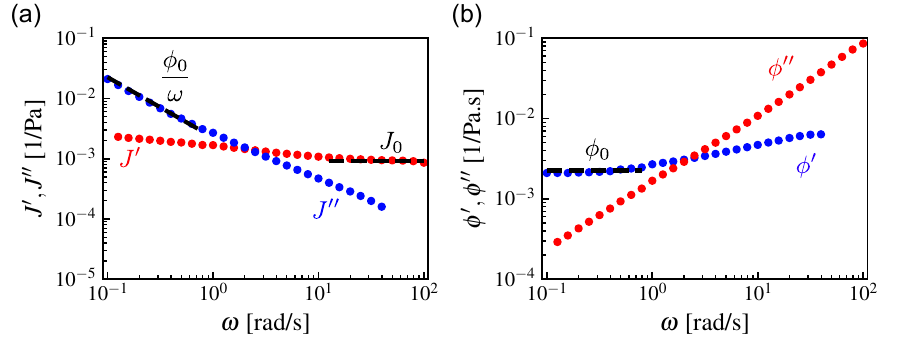}%
 \caption{LVE material functions: (a) elastic compliance $J^{\prime}$ and (b) viscous fluidity $\phi^{\prime}$ show a small change as a function of frequency (both would be constant for a single-mode Maxwell model). The lognormal spectrum fit (Fig.~\ref{LognFit}) is used to compute the elastic plateau modulus $G_N^0=1060$ Pa and average timescale $\tau_0=0.402$ s  from which the elastic compliance $J_0=\frac{1}{G_N^0}=9.43\times 10^{-4}$~$(\mathrm{Pa})^{-1}$ and infinite time fluidity $\phi_0=\frac{1}{G_N^0\tau_0}=2.35\times10^{-3}$~$(\mathrm{Pa.s})^{-1}$ are computed.}%
 \label{fig:3SAOSSweep}
 \end{figure}

The frequency-dependent behavior of the extracted linear elastic compliance $J^{\prime}$ and viscous fluidity $\phi^{\prime}=J^{\prime\prime}\omega$ is presented in Fig.~\ref{fig:3SAOSSweep}. The LVE behavior of PVA-Borax is nearly Maxwellian but deviates from a single-mode Maxwell model due to its distribution of relaxation times, which was captured by a continuous Lognormal distribution (Appendix A). The mean relaxation time $\tau_0=0.402$s, dispersity index \textit{\DJ}$=2.54$, and elastic compliance plateau $J_0=\frac{1}{G_0}=0.9\times 10^{-3}\mathrm{Pa^{-1}}$ were extracted from the moments of the distribution following the work of Martinetti et al\cite{Martinetti2018ContinuousShear}. The relaxation time $\tau_0$ was chosen as the mean relaxation time of the modulus-weighted spectrum $Q(\tau)=\frac{H(\tau)}{\tau}$ (i.e., first-moment average in terms of $G_i$ of a discrete spectrum) and can be related to the plateau modulus $G_0$ and steady-state linear viscosity $\eta_0$ as $\tau_0=(\eta_0-\eta_s)/G_0=\eta_0/G_0$, since the solvent viscosity $\eta_s$ is neglected here. The viscous compliance $J^{\prime\prime}$ at the high frequencies ($\omega>60$ rad/s) is noisy, because it is the sub-dominant compliance component ($\tan \delta<0.1$)\cite{Singh2019OnRepresentation}. This caused the fits to have an unsatisfactory adjusted coefficient of determination $R^2_{adj}$ (documented in the SI, Fig. S1) associated with the goodness of fit, and therefore the corresponding material functions are omitted from the results.

\begin{figure}[h!]
 \includegraphics{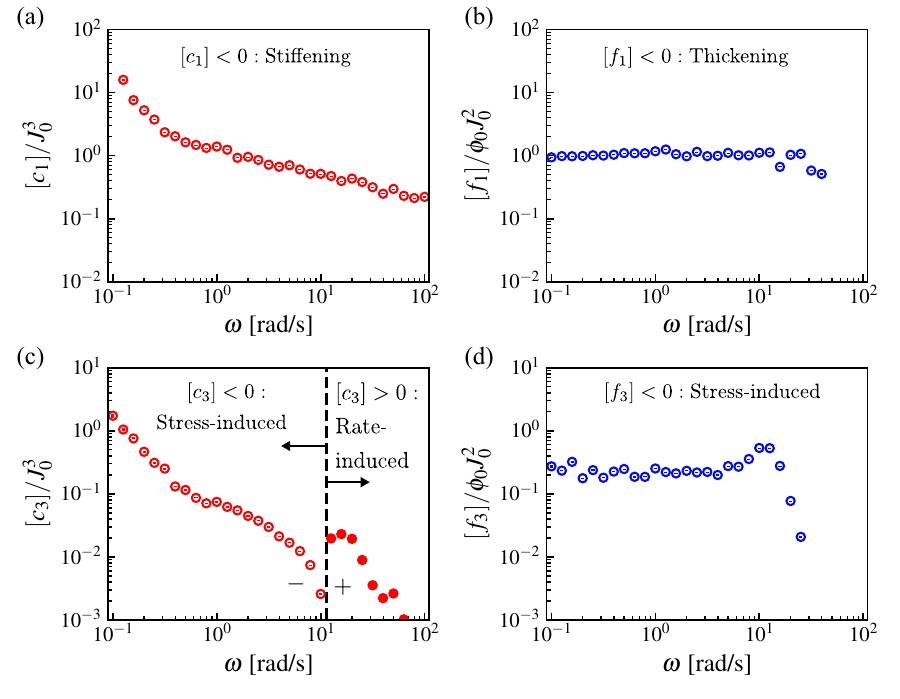}%
 \caption{First-ever MAOStress material functions measurement as a function of frequency. The normalized intrinsic first-harmonic (a) elastic $[c_1]/J_0^3$and (b) viscous $[f_1]/\phi_0 J_0^2$ MAOStress material functions show that PVA-Borax stiffens and thickens under weakly-nonlinear stress forcing. The normalized intrinsic third harmonic (a) elastic $[c_3]/J_0^3$and (b) viscous $[f_3]/\phi_0 J_0^2$ MAOStress material functions indicate the primary forcing mechanism causing the observed stiffening and thickening (either stress or rate-of-stress). Open symbols represent negative values, and closed symbols represent positive values. \label{fig:3MAOSsweep}}%
 \end{figure}

The frequency-dependent MAOStress material functions are reported in Fig.~\ref{fig:3MAOSsweep}(a-d). The sign of the material functions is indicated by whether the symbol is filled (positive) or not (negative). The negative sign of the first-harmonic material functions $[c_1]$ and $[f_1]$ shows that PVA-Borax stiffens and thickens with increasing stress across all the tested frequencies. The value of the normalized material function can be used to compare to models or other material formulations to fit model parameters or establish benchmarks for desired nonlinearity in different applications. The material functions are normalized to represent the relative strength of the nonlinearity to the linear component, which will be further discussed below. The magnitude of $[c_1]$ decreases as a function of frequency, which shows that the stress-stiffening nonlinearity decreases in the elastic limit (high frequencies). On the other hand, the viscous nonlinearity is constant at the lower frequencies until $\omega=10 \;\mathrm{rad/s}$, at which point a sharp decrease is observed, showing that the viscous thickening is weak in the elastic limit. Another reason behind this drop could be the difficulty of measurement because the viscous component is subdominant, as discussed above. The same material's MAOStrain results are unaffected at those frequencies (Fig.~\ref{fig:3MAOSsweep}), which could be due to optimized control for oscillatory strain in commercial instruments compared to oscillatory stress. 
Third harmonics are associated with the distortion of Lissajous curves\cite{Ewoldt2008,Ewoldt2013Low-dimensionalViscoelasticity,Pineiro-Lago2023LargePDO}. The sign of the third harmonics determines the concavity of the strain-stress Lissajous curves (fifth harmonics and higher also contribute, but these are negligible with the weakly nonlinear MAOStress). Third harmonics are best interpreted in the context of the first-harmonic deviations\cite{Ewoldt2013Low-dimensionalViscoelasticity}. A negative value of $[c_3]$ and $[f_3]$ means that within an oscillation cycle in the weakly nonlinear limit, the elastic and viscous compliance of the material decreases with stress and that the degree at which this decrease occurs increases with the stress amplitude (material is stiffer/thicker at higher stress). Combining this intracycle change indicated by $[c_3]$ and $[f_3]$  with the observed intercycle stiffening and thickening at increasing amplitudes, as indicated by the signs of $[c_1]$ and $[f_1]$, we can deduce that the stiffening occurs due to increased stress and not stress rates (following Fig. 7 in Ewoldt and Bharadwaj\cite{Ewoldt2013Low-dimensionalViscoelasticity}). The positive sign of $c_3$ for $\omega>10 \mathrm{rad/s}$ suggests that the stiffening happens in that range primarily due to large stress rates and not stress itself. The ratio of the  $[c_3]$ to $[c_1]$ and $[f_3]$ to $[f_1]$ is comparable to  model predictions from the SSTNM model ($\approx 0.3$) as derived in the SI. More about the mathematics behind the sign and magnitude interpretation of the intrinsic MAOS material functions can be found in prior work from our group\cite{Ewoldt2013Low-dimensionalViscoelasticity}.

 \section{Discussion}
\label{Sec:3Discuss}

\subsection{Normalization of MAOStress functions}
The choice of normalization in Fig.~\ref{fig:3MAOSsweep} was based on the scaling observed in the Strain-Stiffening Transient Network Model (SSTNM)\cite{Bharadwaj2017AShear,Martinetti2018InferringNetwork} and the time-strain separable MAOS model\cite{Martinetti2019Time-strainShear}. Even though they are independent, the linear and nonlinear MAOStress material functions may all scale with a measure of the linear elastic compliance $J_0$ or fluidity $\phi_0$. Similar behavior is observed in MAOStrain, where for many constitutive models, the linear and nonlinear material functions scale with $G_0$ and $\eta_0$. These scalings emerge from the physics that underlies the behavior of the material. The SSTNM model was developed by Bharadwaj et al.\ in 2017 to model the strain-stiffening MAOS behavior of PVA-Borax, where none of the existing nonlinear viscoelastic models was able to capture the behavior of this material\cite{Bharadwaj2017AShear}. 

We solved the SSTNM model numerically in stress control mode (computational details in Appendix B), which yields the normalization used in Fig.~\ref{fig:3MAOSsweep} using $J_0$ and $\phi_0$. The results were corroborated by applying the Lennon et al.\cite{Lennon2020MediumExamples} interrelations to the analytical solution\cite{Bharadwaj2017AShear} of the MAOStrain material functions of the SSTNM model to convert to MAOStress functions.
Moreover, applying the same relations to the time-strain separable MAOS models with a single-mode relaxation kernel yields the same scaling relationship  (derived in Appendix B Eqs.~\ref{eq:3MAOSTSSsol}). The MAOS time-strain separable (TSS) models are a generic category for which time-strain separability holds and where the damping function is expanded up to only the first deviation from linearity\cite{Martinetti2019Time-strainShear}. These findings suggest that the chosen normalization in Fig.~\ref{fig:3MAOSsweep} is suitable for comparing data from different materials as it provides a measure of nonlinearity independent from the linear compliance and fluidity for various models of nonlinear viscoelasticity.

Other choices of MAOStress normalization are possible and can be derived by examining Eqn~\ref{eq:3 harmnonics expansion} but will have different physical interpretations as summarized in Table~\ref{Tab:Interpret}. The unnormalized material functions (e.g., $[c_1]$) quantify the rate at which the absolute nonlinear deviation (e.g., $[c_1]\sigma_0^3$) grows per unit of $\sigma_0^3$. However, when comparing different materials, the ratio of the deviation to the linear term at each frequency may be more relevant. Therefore, another choice to normalize is, e.g., $\frac{[c_1]}{J^{\prime}}(\omega)$ (similar normalization for the other material functions shown in Table~\ref{Tab:Interpret}), which quantifies the rate at which the magnitude of the ratio $\frac{[c_1]\sigma_0^3}{J_1^{\prime}\sigma_0}=\frac{[c_1]\sigma_0^2}{J_1^{\prime}}$ grows per unit of $\sigma_0^2$. Examining the ratio $\frac{[c_1]\sigma_0^2}{J_1^{\prime}}$ provides an interpretation of the scaling inferred from the SSTNM and MAOS TSS models and helps rationalize why $[c_1]$ may scale with $J_0^3$. For a single-mode SSTNM model, $J^{\prime}(\omega)=J_0$, and the stress amplitude $\sigma_0$ is to first-order proportional to $G_0=1/J_0$ for any particular strain. Hence, the relative nonlinearity $\frac{[c_1]\sigma_0^2}{J^{\prime}}$ suggests and justifies the simpler model-informed scaling $[c_1]/J_0^3$, interpreted as a measure of relative nonlinearity at a strain of order one ($\gamma \approx \sigma_0 J^{\prime}= \sigma_0 J_0=1$ in this case). Similar reasoning for fluidity results in, e.g.\ $[f_1]/(J_0^2\phi_0)$, as shown in Table~\ref{Tab:Interpret}. 

The model-informed $[c_1]/J_0^3$ normalization was reasoned from a single relaxation mode. However, PVA-Borax is better described by a distribution of relaxation times, in which case $J_0$ can be replaced with $J^{\prime}$ and $\phi_0$ with $\phi^{\prime}$ (as in Table~\ref{Tab:Interpret} (multi-mode)) to have an analogous interpretation at different frequencies. Comparing the frequency dependence of $\frac{[c_1]}{J_0^3}$ and $\frac{[c_1]}{J^{\prime 3}}$ in Fig.~\ref{fig:3MAOSsweepnorm2} shows that the latter normalization varies less with frequency achieving a constant plateau for $\omega>0.3$ rad/s. This observation indicates that incorporating the frequency dependence of the linear material functions in the normalization helps reduce the complexity of the MAOStress description, which shows a near-constant relative nonlinearity as a function of frequency. This observation is remarkable and is associated with a weak frequency dependence of the critical stress amplitude, as we explore in detail below.

\begin{table}
\caption{Normalization choices for MAOStress intrinsic nonlinearities. Frequency dependence is implied for all variables except the constants $J_0$ and $\phi_0$. 
 \label{Tab:Interpret}}
\begin{center}
\renewcommand\arraystretch{1.2}
\begin{tabular}{| >{\raggedright\arraybackslash}p{0.33\linewidth} | >{\raggedright\arraybackslash}p{0.33\linewidth} |>{\raggedright\arraybackslash}p{0.33\linewidth}|}
\hline
\textbf{Type} & \textbf{Material Functions} & \textbf{Comments/Interpretation} \\ \hline
Absolute nonlinearities &        $[c_1]$, $[f_1]$, $[c_3]$, $[f_3]$            &        Rate of growth of harmonic deviation per $\sigma_0^3$   \\ \hline
   Relative nonlinearities  &        $\frac{[c_1]}{J^{\prime}}$, $\frac{[f_1]}{\phi^{\prime\prime}}$, $\frac{[c_3]}{J^{\prime}}$, $\frac{[f_3]}{\phi^{\prime\prime}}$            &   Rate of growth of relative harmonic deviation per $\sigma_0^2$          \\\hline
Model-informed normalized nonlinearity (single-mode)& $\frac{[c_1]}{J_0^3}$, $\frac{[f_1]}{J^2_0 \phi_0}$, $\frac{[c_3]}{J_0^3}$, $\frac{[f_3]}{J^2_0 \phi_0}$    & Dimensionless. Quantifies relative nonlinearity at $\gamma \approx J_0\sigma_0= 1$    \\ \hline
  Model-informed normalized nonlinearity (multi-mode)& $\frac{[c_1]}{J^{\prime 3}}$, $\frac{[f_1]}{J^{\prime 2} \phi^{\prime}}$, $\frac{[c_3]}{J^{\prime 3}}$, $\frac{[f_3]}{J^{\prime 2} \phi^{\prime}}$    & Accounts for multi-mode relaxation (varying $J^{\prime}(\omega)$ and $\phi^{\prime}(\omega)$) 
\\ \hline

\end{tabular}

\end{center}
 
\end{table}

\subsection{Defining a threshold-independent critical forcing amplitude}

The emergence of nonlinearities is gradual, and the choice of where the linear regime ends is arbitrary, e.g., depending on a specified threshold, say, $\epsilon$. For example, Pearson and Rochehfort reported one of the earliest experimental measurements of a frequency-dependent critical strain $\gamma_{crit}(\omega)$ in 1982\cite{Pearson1982BehaviorFields}, using the deviation of first-harmonic oscillatory measures. They defined the elastic critical strain at $\big|\frac{G^{\prime}_{1}-G^{\prime}}{G^{\prime}}\big|=5\%$ and the viscous critical strain at $\big|\frac{G^{\prime\prime}_{1}-G^{\prime\prime}}{G^{\prime\prime}}\big|=5\%$, where $G^{\prime}_{1}=\sigma^{\prime}_{1}/\gamma_0$ and $G^{\prime\prime}_{1}=\sigma^{\prime\prime}_{1}/\gamma_0$ (the stress harmonics $\sigma^{\prime}_1$ and $\sigma^{\prime\prime}_1$ are defined in Appendix C below).  
The resulting values of $\gamma_{crit}$ naturally depend on the choice of $\epsilon$, and therefore the shape of $\gamma_{crit}(\omega)$ may also depend on the chosen $\epsilon$. 

MAOS enables a rigorous and threshold-independent study of the frequency dependence of the linearity limits $\sigma_{crit}(\omega)$, $\gamma_{crit}(\omega)$, and $\dot{\gamma}_{crit}(\omega)$.
Ewoldt and Bharadwaj used MAOStrain material functions as a rigorous way to define linearity limit lines \cite{Ewoldt2013Low-dimensionalViscoelasticity} where the choice of $\epsilon$ is transparent. MAOS linearity limits directly compare the weakly nonlinear terms in the expansion to linear terms, providing a quantitative measure of nonlinear emergence and, as we show here, a critical forcing strength (strain, strain-rate, or stress) can be defined independent of the chosen threshold. 

The functions $\sigma_{crit}(\omega)$, $\gamma_{crit}(\omega)$, and $\dot{\gamma}_{crit}(\omega)$ can be interpreted as regime boundaries in Pipkin maps. 
The Pipkin map was initially proposed in 1972 as a tool to map the behavior of complex fluids based on two key parameters: timescale (frequency) and forcing amplitude \cite{Pipkin1986LecturesTheory}. The timescale dependence in a Pipkin map is commonly captured by the Deborah number $\mathrm{De} \equiv \omega \tau$ or $\mathrm{De} \equiv \tau/t$, where $\tau$ is a characteristic relaxation time of the material, $\omega$ is the forcing frequency, and $t$ is a characteristic time scale of the forcing. The Pipkin map has proven to be a valuable means of mapping rheometric conditions to application conditions, identifying regimes of applicability of constitutive models, and comparing the behavior of complex fluids~\cite{Corman2019MappingIntuition,EwoldtMcKinley2017,Macosko1994Rheology:Applications}. Notably, Pipkin did not specify a particular measure of the forcing amplitude, leaving it open for interpretation. Commonly used measures include strain amplitude and the strain-rate amplitude (or Weissenberg number, sometimes written as Wi$\equiv \dot{\gamma\tau}$)\cite{Macosko1994Rheology:Applications,Dealy1999MeltProcessing,Ewoldt2013Low-dimensionalViscoelasticity,Giacomin2011Large-amplitudeModel}. 
In the case of strain amplitude, $\gamma_{crit}(\mathrm{De})$, the linearity limit line for a viscoelastic fluid exhibits a critical strain that diverges as the Deborah number (De) approaches zero according to the general theory of simple fluids by Coleman and Noll\cite{Coleman1961FoundationsViscoelasticity,Coleman1964Erratum:Viscoelasticity}. Conversely, when using strain rate or the Weissenberg number, $Wi_{crit}(\mathrm{De})$, the linearity limit line tends to approach a constant value in the viscous limit and diverges to infinity in the elastic limit.


The frequency-dependent linearity limit based on stress, $\sigma_{cirt}(\omega)$, has not been previously predicted or measured, and MAOStress offers an effective means of experimental and numerical exploration. In this work, we use the MAOStress material functions to compute critical stresses for nonlinearity $\sigma_{crit}$ based on the four independent material functions. Setting the ratio of the elastic and viscous weakly nonlinear deviations $[c_1]\sigma_0^3$ and $\frac{[f_1]}{\omega}\sigma_0^3$ to the corresponding linear terms $J^{\prime}\sigma_0$ and $J^{\prime\prime}\sigma_0$ to an as-yet-unspecified threshold $\epsilon$, we can define critical stresses based on the first-harmonic nonlinearities as:
\begin{subequations}
\label{eq:3critStressFirstHarmonics}
\begin{eqnarray}
\frac{[c_1]\sigma_{crit}^3}{J^{\prime}\sigma_{crit}}= \epsilon \qquad \Rightarrow \qquad \frac{\sigma_{crit,[c_1]}(\omega)}{\sqrt{\epsilon}}=\sqrt{\frac{ J^{\prime}}{[c_1]}}\\
\frac{[f_1]\sigma_{crit}^3}{J^{\prime\prime}\omega\sigma_{crit}}=\epsilon \qquad \Rightarrow \qquad \frac{\sigma_{crit,[f_1]}(\omega)}{\sqrt{\epsilon}}=\sqrt{\frac{ \phi^{\prime}}{[f_1]}}.
\end{eqnarray}
\end{subequations}
Similarly, for the third harmonics, we set the ratio of the elastic and viscous weakly nonlinear deviations $[c_3]\sigma_0^3$ and $\frac{[f_3]}{3 \omega}\sigma_0^3$ to the corresponding linear terms $J^{\prime}\sigma_0$ and $J^{\prime\prime}\sigma_0$ to the threshold $\epsilon$ to obtain:
\begin{subequations}
\label{eq:3critStressThirdHarmonics}
\begin{eqnarray}
\frac{[c_3]\sigma_{crit}^3}{J^{\prime}\sigma_{crit}}=\epsilon \qquad &\Rightarrow \qquad \frac{\sigma_{crit,[c_3]}(\omega)}{\sqrt{\epsilon}}=\sqrt{\frac{ J^{\prime}}{[c_3]}}\\
\frac{[f_3]\sigma_{crit}^3}{3 J^{\prime\prime}\omega\sigma_{crit}}=\epsilon \qquad &\Rightarrow  \qquad \frac{\sigma_{crit,[f_3]}(\omega)}{\sqrt{\epsilon}}=\sqrt{\frac{3  \phi^{\prime}}{[f_3]}}.
\end{eqnarray}
\end{subequations}
Note that we keep $\epsilon$ on the left-hand side of Eqs.~\ref{eq:3critStressFirstHarmonics}--\ref{eq:3critStressThirdHarmonics}, thereby creating new definitions for critical forcing amplitudes that incorporate arbitrary threshold $\epsilon$. That is, the threshold $\epsilon$ can be unspecified (since it can be arbitrarily large or small, based on the intention of application), allowing for comparison of critical forcing strengths independent of the choice of $\epsilon$. For example, choosing $\epsilon=10\%$ rather than $\epsilon=0.1\%$ will increase the critical stress by 10. However, reporting the ratio $\sigma_{crit}/\sqrt{\epsilon}$ will allow for comparison between different works, where a specific critical value can be calculated for any threshold desired. This applies to other measures of nonlinearity, such as strain and strain rate. 

Therefore, we recommend, in general, that future works consider reporting the threshold-normalized critical forcing amplitudes, $[\sigma_{crit}]$, $[\gamma_{crit}]$, and $[\dot{\gamma}_{crit}]$ as 
\begin{equation}\label{eq:eps-norm-amplitudes}
  [\sigma_{crit}]\equiv\frac{\sigma_{crit}}{\sqrt{\epsilon}},\qquad 
  [\gamma_{crit}]\equiv\frac{\gamma_{crit}}{\sqrt{\epsilon}},\qquad
  [\dot{\gamma}_{crit}]\equiv\frac{\dot{\gamma}_{crit}}{\sqrt{\epsilon}}.  
\end{equation}
This normalization allows for better comparison of critical forcing amplitudes, but we acknowledge that it strictly applies for asymptotic deviation from the linear limit theory where higher order terms can be neglected. 

\begin{figure}[h!]
\centering
 \includegraphics{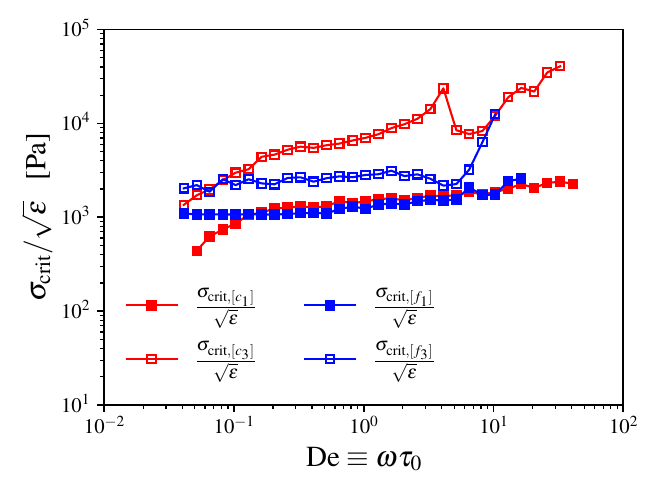}%
 \caption{The theshold-independent critical stress for nonlinearity for each of the four independent MAOStress signals. The critical stress for nonlinearity was computed based on each of the MAOStress material functions, reported in Fig.~\ref{fig:3MAOSsweep}, using Eqs.~\ref{eq:3critStressFirstHarmonics}(a-b) and ~\ref{eq:3critStressThirdHarmonics}(a-b). The elastic (red) and viscous (blue) material functions reveal comparable critical stress for the first harmonics but differ slightly for the third harmonics.    \label{fig:3critStressExpAll}}%
 \end{figure}

Figure~\ref{fig:3critStressExpAll} shows how each critical stress depends on Deborah number by applying Eqs.\ref{eq:3critStressFirstHarmonics}--\ref{eq:3critStressThirdHarmonics} to the data in Fig.~\ref{fig:3MAOSsweep}. The resulting lines represent potential boundaries of the linear regime on a Pipkin map constructed using stress to measure the nonlinearity strength. Stress values below the line are within the linear regime to within a deviation threshold $\epsilon$ of each MAOStress metric. The results show that the MAOStress first-harmonic has the lowest critical stress and is, therefore, the first to reach a nonlinearity threshold, also observed in the amplitude sweeps shown in Fig.~\ref{fig:3AmpSweep}. Moreover, the first harmonics have a weak frequency dependence compared to the more complex dependence exhibited by strain and strain rate first harmonics, as shown in Appendix C (Fig.~\ref{fig:3MAOStrainFirstAndThird}). A peak occurs at 10 rad/s for the third harmonics, concurrent with the sign change in $[c_3]$ and drop in $[f_3]$. The results in Fig.~\ref{fig:3critStressExpAll} reveal that the MAOStress behavior of this PVA-Borax formulation was accessible because the stress required to measure the nonlinear material functions was in an accessible range of stress within the limits of the instrument (and that did not cause irreversible changes to the material). 

\subsection{Pipkin Map linearity limits: Stress, Strain, or Strain-rate}
This section will compare the Pipkin map regime boundary of the linear regime when either stress, strain, or strain rate is considered. To facilitate this comparison, we derived a single critical stress instead of four values from the different material functions. One viable approach is to choose the minimum of the four independent critical stresses defined in Eqs.~\ref{eq:3critStressFirstHarmonics} and ~\ref{eq:3critStressThirdHarmonics}. The drawback of this approach is that the critical stress linearity limit line will be noisy as the criteria shift between different metrics. To resolve this issue, we define a single critical stress based on the first harmonics by finding the stress at which the magnitude of the total first-harmonic deviation exceeds a threshold in comparison to the magnitude of the linear first-harmonic complex compliance
\begin{equation}
\frac{\sqrt{([c_1]\sigma_{crit}^3)^2+(\frac{[f_1]}{\omega}\sigma_{crit}^3)^2}}{\sqrt{(J^{\prime}\sigma_{crit})^2+(J^{\prime\prime}\sigma_{crit})^2}}=\epsilon \qquad \Rightarrow \qquad \frac{\sigma_{crit,1}(\omega)}{\sqrt{\epsilon}}=\sqrt{\frac{|J^{*}|}{\sqrt{([c_1])^2+\left(\frac{[f_1]}{\omega}\right)^2}}}.
\label{eq:3pipkinLimitsMAOStressFirst}
\end{equation}
$\sigma_{crit,1}$ is the stress at which the first-harmonic deviations affect both the observed elastic and viscous Lissajous curves, particularly in terms of the rotation and enclosed area. Moreover, we define $\sigma_{crit,3}$ as the critical stress at which the magnitude of the total third harmonic deviation exceeds a threshold with respect to the magnitude of the to the magnitude of the linear first-harmonic complex compliance
\begin{equation}
\label{eq:3pipkinLimitsMAOStressThird}
\frac{\sqrt{([c_3]\sigma_{crit}^3)^2+(\frac{[f_3]}{3\omega}\sigma_{crit}^3)^2}}{\sqrt{(J^{\prime}\sigma_{crit})^2+(J^{\prime\prime}\sigma_{crit})^2}}=\epsilon  \qquad \Rightarrow \qquad  \frac{\sigma_{crit,3}(\omega)}{\sqrt{\epsilon}}=\sqrt{\frac{ |J^{*}|}{\sqrt{([c_3])^2+\left(\frac{[f_3]}{3\omega}\right)^2}}}.
\end{equation}
$\sigma_{crit,3}$ is the stress at which the third-harmonic deviations affect both the observed elastic and viscous Lissajous curves, particularly in terms of the distortion from their linear elliptical shape. Finally we define $\sigma_{crit}$ as the minimum of $\sigma_{crit,1}$ and $\sigma_{crit,3}$:
\begin{equation}
    \sigma_{crit}=\mathrm{min}(\sigma_{crit,1},\sigma_{crit,3}).
    \label{Eq:3criticalStressFinal}
\end{equation}

Similarly, the critical strain for nonlinearity can be computed from the MAOStrain material functions, measured on the ARES-G2 rheometer as described in Appendix C. The critical strain $\gamma_{crit}$ is the minimum of $\gamma_{crit,1}$ and $\gamma_{crit,3}$
\begin{equation}
        \gamma_{crit}=\mathrm{min}(\gamma_{crit,1},\gamma_{crit,3}),
    \label{Eq:3criticalStrainFinal}
\end{equation}
which are defined as
\begin{equation}
\label{eq:3critStrainFirstHarmonics}
\frac{\sqrt{([e_1]\gamma_{crit}^3)^2+([\omega v_1]\gamma_{crit}^3)^2}}{\sqrt{(G^{\prime}\sigma_{crit})^2+(G^{\prime\prime}\sigma_{crit})^2}}=\epsilon \qquad \Rightarrow \qquad \frac{\gamma_{crit,1}(\omega)}{\sqrt{\epsilon}}=\sqrt{\frac{ |G^{*}|}{\sqrt{([e_1])^2+(\omega[v_1])^2}}},
\end{equation}
and
\begin{equation}
\label{eq:3critStrainThirdHarmonics}
\frac{\sqrt{([e_3]\gamma_{crit}^3)^2+([\omega v_3]\gamma_{crit}^3)^2}}{\sqrt{(G^{\prime}\sigma_{crit})^2+(G^{\prime\prime}\sigma_{crit})^2}}=\epsilon\qquad  \Rightarrow \qquad  \frac{\gamma_{crit,3}(\omega)}{\sqrt{\epsilon}}=\sqrt{\frac{ |G^{*}|}{\sqrt{([e_3])^2+(\omega[v_3])^2}}}.
\end{equation}
The critical strain-rate amplitude can be obtained from the critical strain as \begin{equation}
    \dot{\gamma}_{crit}=\gamma_{crit}\omega.
\end{equation}

\begin{figure}[h!]
 \centering
 \includegraphics{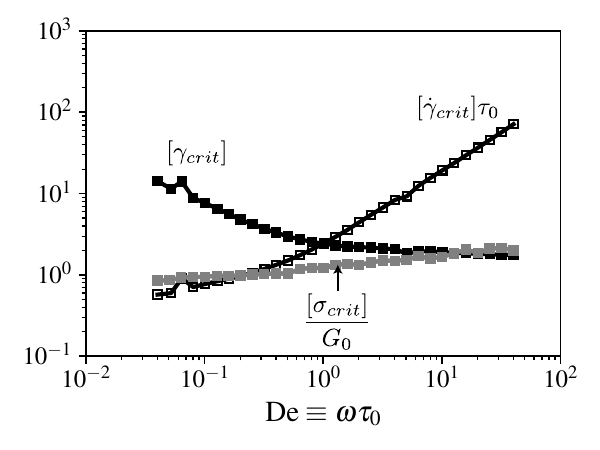}%
 \caption{Comparing Pipkin Diagram nonlinear transition limits. The threshold-independent critical stress, strain, and strain rate values estimated from MAOStrain (Eq.\ref{Eq:3criticalStressFinal})and MAOStress (Eq.\ref{Eq:3criticalStrainFinal}) material functions (Figs. \ref{fig:3SAOSSweep} and \ref{fig:3MAOSsweep}) obtained experimentally for PVA-Borax 1.25/2.75 wt\%. The critical stress appears to be a more universal measure of nonlinearity as a function of frequency, where it varies between the finite critical strain rate and strain limits at low and high De, respectively.\label{fig:3pipkinMapExp}}%
 \end{figure}

Figure~\ref{fig:3pipkinMapExp} compares the De-dependent critical stress, strain, and strain rate, all dimensionless (as $\sigma_{crit}/G_0$, $\gamma_{crit}$, and $\tau_0\dot\gamma_{crit}$) and normalized by $\sqrt{\epsilon}$ following Eq.~\ref{eq:eps-norm-amplitudes}. Remarkably, the dimensionless critical stress approaches equivalence with the critical strain in the elastic limit (high Deborah number) and approaches the dimensionless critical strain rate in the viscous limit (low Deborah number). 

We hypothesize that this result can be justified by first considering that the stress is dominated by the viscous contribution at low frequencies and elastic effects at high frequencies for a viscoelastic liquid. Specifically for PVA-Borax, the emergence of nonlinearity is directly related to the local stretch (deformation) of the polymer chains\cite{Bharadwaj2017AShear,Martinetti2018InferringNetwork}, rather than other deviations from equilibrium such as polymer orientation or network junction breakage. Therefore, we can hypothesize that nonlinearity arises due to the micromechanical strain of elastic (polymeric) units.  

In the elastic limit $\mathrm{De}\to \infty$, the deformation timescales are shorter than the timescale of transient network bond dissociation, and the applied strain is fully translated into the micromechanical stretch of the polymer strands, which will exceed the linearity threshold at a critical characteristic forcing strain $\gamma_{crit}$, consistent with the physical picture. Quantitatively, this critical strain maps directly to a critical elastic stress $\sigma_{crit}=G_0 \gamma_{crit}$, equal to the critical total stress if purely viscous solvent contributions are neglected and if strain nonlinearities are neglected. In the absence of purely viscous stresses, this predicts $\gamma_{crit} = \sigma_{crit}/G_0$ in the limit $\mathrm{De} \rightarrow \infty$, a close correspondence that we observe in Fig.~\ref{fig:3pipkinMapExp}. We specifically observe a slightly larger dimensionless critical stress, consistent with a purely viscous solvent being present that adds to the total observed stress. Another explanation of the slight discrepancy is that the approximation $\sigma_crit=G_0\gamma_{crit}$ will underpredict the critical stress since, for a stiffening material like PVA-Borax, the effective modulus will be higher than the linear plateau modulus $G_0$. Therefore, the dimensionless critical stress calculated as  $\sigma_{crit}/G_0$ will be higher than the critical strain $\gamma_{crit}$.

In the viscous limit $\mathrm{De} \to 0$, the stress is independent of the total accumulated strain and instead is proportional to the strain rate. 
We expect a critical strain rate $\dot{\gamma}_{crit}$ associated with a critical stress as $\sigma_{crit} \approx \eta_0\dot{\gamma}_{crit}$ (neglecting solvent viscosity and nonlinearity of the stress-strain rate relationship).
The nonlinearity for PVA-Borax is caused by micromechanical polymer strain, which in this limit is proportional to the applied strain rate. 
The critical dimensionless stress is then $\sigma_{crit}/G_0 = (\eta_0/G_0)\dot\gamma_{crit}$, or in terms of dimensionless critical strain rate, $\sigma_{crit}/G_0 = \tau_0\dot\gamma_{crit}$ 
This rationalizes the close correspondence of dimensionless critical stress and strain rate in the limit $\mathrm{De} \rightarrow 0$, as we observe in Fig.~\ref{fig:3pipkinMapExp}. The slightly higher normalized critical stress, however, can also be explained by the neglected solvent viscosity and thickening nonlinearity of PVA-Borax in the prediction $\sigma_{crit}\approx \eta_0 \dot{\gamma}_{crit}$. 

The critical stress values at $\mathrm{De} \ll 1$ and $\mathrm{De} \gg 1$ are similar but different, and some insight is provided by considering the critical micromechanical strain (polymer stretch) for PVA-Borax in steady flow. 
The accumulated polymer strain under steady flow is $\propto \dot{\gamma}\tau$ (equivalent to a Weissenberg number\cite{White1964DynamicsSpinning}), where $\tau$ is a characteristic relaxation time of the elastic components that determines the accumulated micromechanical strain for a given strain rate $\dot{\gamma}$. Assuming that the critical accumulated micromechanical strain is proportional to $\gamma_{crit}$ observed in the limit $\mathrm{De} \gg 1$, this predicts $\dot\gamma_{crit} \approx \gamma_{crit}/\tau$.
Then in the $\mathrm{De} \ll 1$ limit, the critical stress is $\sigma_{crit}=\eta_0\dot{\gamma_{crit}}$, or $\sigma_{crit}/G_0=(\tau_0/\tau)\gamma_{crit}$. 
Therefore, the ratio $\tau_0/\tau$ may rationalize the difference in critical stress in the limit $\mathrm{De} \ll 1$ compared to $\mathrm{De} \gg 1$. 
We observe a lower critical stress for $\mathrm{De} \ll 1$, suggesting that the ratio $\tau_0/\tau < 1$. 
If the characteristic timescale $\tau$ that determines the accumulated critical polymer stretch in steady flow is close to the average relaxation time $\tau_0$, then the critical stress in the elastic limit and viscous limit will be similar. 
However, if these two timescales differ (e.g.\ for polydisperse systems), then the critical stress may be different. For example, Nichetti and Manas-Zloczower derived that for a polymer melt with a lognormal distribution of molecular weights, where the critical stress is constant for all molecular weights, the 

The potential universality of critical stress might extend to a broad range of viscoelastic liquids. It is conceivable that any given viscoelastic liquid could exhibit a finite critical strain, $\gamma_{crit}$, in the elastic limit ($\mathrm{De}\to\infty$) and a finite critical strain rate, $\dot{\gamma}_{crit}$, in the viscous limit ($\mathrm{De} \to \infty$). This would imply that the critical stress ($\sigma_{crit}= G_0\gamma_{crit}$ for $\mathrm{De} \gg 1$ and $\sigma_{crit}= \eta_0 \dot{\gamma}_{crit}$ for $\mathrm{De} \ll 1$) approach plateaus in these limits. However, the critical strain and strain rate are expected to exhibit diverging behavior in the limit of $\mathrm{De} \ll 1$ and $\mathrm{De} \gg 1$, respectively. Therefore, while our experimental observations suggest the universality of critical stress will hold more generally, further research is needed to confirm and generalize this finding to other materials. Supporting evidence already exists. For example, our observations align with an earlier study by Hatzikiriakos and Dealy \cite{Hatzikiriakos1991WallStudies}, which reported a frequency-independent critical stress for instability/wall slip in a sliding plate rheometer, another type of nonlinearity under oscillatory shear deformations.

The universality of critical stress may allow for faster characterization of material properties using the frequency-sweep MAOS method\cite{Singh2017Frequency-sweepMAOS}. This technique eliminates the need for a full amplitude sweep at every frequency and instead relies on the ability to predict at which amplitude the MAOS material functions can be measured. Based on that prediction, a single amplitude measurement is done at every frequency. Since the critical stress varies slightly as a function of De compared to critical strain, applying the frequency-sweep MAOS technique in MAOStress should be simpler. This will result in a high-throughput characterization while minimizing the impact of chemical and physical changes in the loaded material. A similar phenomenon is utilized in medium amplitude superposition (MAPS) measurements, where the optimal window is frequency-independent in stress-controlled mode\cite{Lennon2020MediumAnalysis}. Additionally, this work shows that the change in the critical stress amplitude required to measure MAOStress nonlinearities is much smaller than MAOStrain measurements, making the MAOStress technique potentially more accessible than MAOStrain. 
\section{Conclusion}
The MAOStress response of PVA-Borax was thoroughly studied, showing that stress-controlled measurements can offer another route for testing the weakly nonlinear regime and answer fundamental questions regarding the origins of nonlinearity in viscoelastic materials. The MAOStress results confirmed the stiffening nature of PVA-Borax and showed that stress is a more universal measure of nonlinearity strength across a wide range of De, consistent with the microstructure of PVA-Borax. Moreover, the measured linearity limits can be interpreted as Pipkin map regime boundaries, offering experimental testing and validation of theoretical predictions. The SSTNM model was used to derive a viable normalization of the nonlinear MAOStress material functions that account for their scaling with the linear material properties. The frequency dependence of the normalized material functions revealed a simplicity in the features of MAOStress material functions compared to MAOStrain. This simplicity could be harnessed in future works that compare MAOStress measurements with model predictions for physical inference.

The intrinsic nonlinearities were used to derive a threshold-independent critical forcing amplitude and to show that stress is a more universal measure of nonlinearity strength as its frequency variability at the emergence of nonlinearity is much smaller than MAOStrain. This fundamental result should be considered when studying deformation-induced physical phenomena at different timescales or frequencies. Moreover, this result has significant implications for small amplitude oscillatory shear measurements (SAOS). A common practice for choosing a strain or stress amplitude for a linear SAOS measurement is to do an amplitude sweep at different frequencies and choose a strain or stress in different frequency ranges with a good signal-to-noise ratio without being in the nonlinear regime. With the insights from section IV, it may suffice to only do an amplitude sweep at each end of the frequency range to find a suitable stress amplitude for the entire frequency range of a measurement. 

The findings of this work motivate future exploration of the MAOStress response of other materials. For example, comparing the MAOStress material functions of multiple PVA-Borax formulations could be very insightful in understanding the effect of microstructural changes on the material properties. Moreover, testing different viscoelastic materials, such as melts, colloids, and multi-component systems, will also highlight the effect of the microstructure on the MAOStress response.

\section*{Acknowledgements}
This work was funded by the U.S. Department of Energy, Office of Basic Energy Sciences, Division of Materials Sciences and Engineering under Award DE-SC0020858, through the Materials Research Laboratory at the University of Illinois at Urbana-Champaign. The authors are grateful to Anton Paar for loaning the MCR 702 instrument to the Ewoldt Research Group, which enabled the MAOStress measurements.
\section*{Data Availability}
Data available on request from the authors.

\setcounter{table}{0}
\renewcommand{\thetable}{A\arabic{table}}%
\setcounter{figure}{0}
\renewcommand{\thefigure}{A\arabic{figure}}
\section*{Appendix A: Identifying lognormal spectrum parameters}

The linear viscoelastic MAOStress material functions in Figure \ref{fig:3SAOSSweep} were fit with the lognormal relaxation spectrum. The lognormal spectrum is a parametrization for the distribution of relaxation timescales given as
\begin{equation}
    H(\tau)=H_{\mathrm{max}}e^{-\frac{1}{2}(\frac{\log \tau -\log \tau_{max}}{\hat{\sigma}})^2}
\end{equation}
where $H_{\mathrm{max}}$ defines the spectrum peak, $\tau_{\mathrm{max}}$ defines the location of the peak, and $\hat{\sigma}$ defines the breadth of the distribution. The linear moduli can be computed directly from the spectrum at every frequency as
\begin{equation}
    G^{\prime}(\omega)=\int_0^{\infty}\frac{H(\tau)}{\tau}\frac{(\omega\tau)^2}{1+(\omega\tau)^2}d\tau,
\end{equation}
and 
\begin{equation}
    G^{\prime\prime}(\omega)=\int_0^{\infty}\frac{H(\tau)}{\tau}\frac{\omega\tau}{1+(\omega\tau)^2}d\tau.
\end{equation}
The linear compliances $J^{\prime}$ and $J^{\prime\prime}$ are computed using the linear regime interchangeability relations \begin{equation}
J^{\prime}=\frac{G^{\prime}}{G^{\prime 2}+G^{\prime\prime 2} },
\label{eq:3LinInter1}
\end{equation}
and 
\begin{equation}
J^{\prime\prime}=\frac{G^{\prime\prime}}{G^{\prime 2}+G^{\prime\prime 2} }.
\label{eq:3LinInter2}
\end{equation}
The algorithm finds the parameter set $\textbf{p}=$[$H_{max}$, $\tau_{max}$, $\hat{\sigma}$] that minimizes the residual error between $J^{\prime}(\omega)$ and $J^{\prime \prime}(\omega)$ measured experimentally and those computed from the model. The data representation and definition of residual are important choices when fitting~\cite{Singh2019OnRepresentation}. The residual error here was defined as \begin{equation}
    \mathrm{Res}=\sum_{i=1}^{N_d}((\log J^{\prime }_D(\omega_i)-\log J^{\prime }_M(\omega_i;\textbf{p}))^2+(\log J^{\prime\prime }_D(\omega_i)-\log J^{\prime\prime }_M(\omega_i;\textbf{p}))^2),
\end{equation}
where $\{J^{\prime}_D, J^{\prime\prime}_D\}$ and $\{J^{\prime}_M, J^{\prime\prime}_M\}$ are the compliances from the data and model, respectively. 

\begin{figure}
 \includegraphics{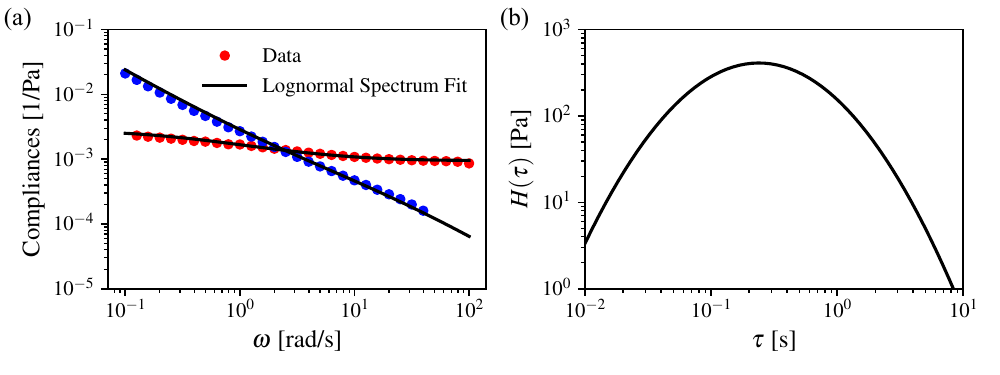}%
 \caption{ Lognormal spectrum LVE fit. The lognormal spectrum is used to fit the (a) LVE frequency sweep of PVA-Borax to extract an average time scale $\tau_0=0.404$ s, plateau modulus $G_0=1040$ Pa, and dispersity \textit{\DJ}$=2.655$. The best fit model parameters are $H_{max}=410$ Pa, $\tau_{max}=0.24$ s, and $\hat{\sigma}=0.988$. The corresponding relaxation spectrum is shown in (b)}\label{LognFit}%
 \end{figure}

The lognormal spectrum can be used to compute an average relaxation time ($\tau_0=\tau_{max}e^{\frac{1}{2}\hat{\sigma}^2}$), plateau modulus $G_0=H_{max}\hat{\sigma}\sqrt{2\pi}$, and dispersity \textit{\DJ}$=e^{\hat{\sigma}^2}$, based on Martinetti et al\cite{Martinetti2018ContinuousShear}. The values for the PVA-Borax sample tested are reported in Figure \ref{LognFit}.

\section*{Appendix B: Model-informed MAOStress Material Functions Normalization}
The intrinsic MAOStress material functions $[c_1]$, $[f_1]$, $[c_3]$, and $[f_3]$ of PVA-Borax are in the order of $~10^{-9}-10^{-12}$ $\mathrm{Pa^{-3}}$ for elastic nonlinearities and$~10^{-9}-10^{-12}$ $\mathrm{Pa^{-3}s^{-1}}$ for the viscous nonlinearities. Both the units and the scale motivate the search for a normalization that accounts for the scaling of these nonlinear material functions with the linear material functions and that provides a measure of the relative nonlinearity that allows for comparison with other materials of interest. Hence, we solved the SSTNM model in the MAOStress regime to explore this scaling. First, the SSTNM model is solved numerically, which reveals that the material functions scale according to the normalization shown in Figure \ref{fig:3MAOSsweep}. Second, we use the Lennon et al. interrelations\cite{Lennon2020MediumExamples} to find the exact analytical solution of MAOStress material functions for the SSTNM model. The results show that the numerical and analytical solutions match and validate each other. Finally, we use the validated Lennon et al. interrelations to find the analytical solution of the general MAOS TSS model, which exhibited the same scaling as the SSTNM model. 
\subsection*{Numerical Solution of SSTNM model in MAOStress}
To solve the SSTNM model numerically, we start with the differential equations given by Bharadwaj et al.\cite{Bharadwaj2017AShear}(their sections IV.A and IV.B.). The stress tensor $\boldsymbol{\sigma}$ is written in terms of a stretch vector $\underline{Q}$ and a microstructural conformation tensor $\textbf{A}=\frac{<\underline{Q}\underline{Q}>}{Q_{eq,i}^2}$ as 
\begin{equation}
\label{eq:3StressTensorSSTNM}
    \boldsymbol{\sigma}=H(Q)Q_{eq,i}^2 \textbf{A},
\end{equation}
where $Q_{eq,i}$ is the $i^{th}$ component of the equilibrium end-to-end vector between neighbouring network junctions at equilibrium, and $Q=Q_{eq,i}tr(\textbf{A})$ is the average stretch magnitude. The time evolution of  $\textbf{A}$ is given by
\begin{equation}
\stackrel{\nabla}{\mathbf{A}}=-\frac{1}{\tau_0}(\textbf{A}-\textbf{I}),
\label{Eqn:AEvolution}
\end{equation}
where $\tau_0$ is the relaxation timescale, $\mathbf{I}$ is the identity tensor, and $\stackrel{\nabla}{\mathbf{A}}$ is the upper convected time derivative. This derivative is frame-indifferent and is given by
\begin{equation}
    \stackrel{\nabla}{\mathbf{A}}\equiv 
    \frac{D\mathbf{A}}{Dt}-(\underline{\nabla}\,\underline{v})^T\cdot\mathbf{A}-\mathbf{A}\cdot(\underline{\nabla}\,\underline{v}),
\end{equation} where $\underline{v}$ is the local velocity vector and $\frac{D\mathbf{A}}{Dt}$ is the material derivative given by $\frac{D\mathbf{A}}{Dt}=\frac{\partial \mathbf{A}}{\partial t}+(\underline{v}\cdot\underline{\nabla})\mathbf{A}$. For simple shear flow and assuming $A_{ii}(t=0)=1$, Eq. \ref{Eqn:AEvolution} simplifies to two differential equations
\begin{subequations}
\label{eq:3A-Evolution-set}
\begin{eqnarray}
\frac{dA_{11}}{dt}-2\dot{\gamma}A_{21}=-\frac{1}{\tau_0}(A_{11}-1),\\
\frac{dA_{21}}{dt}-2\dot{\gamma}=-\frac{1}{\tau_0}(A_{21}).
\end{eqnarray}
\end{subequations}
We assume a simplified finitely extensible nonlinear elastic (FENE) model functional form for the nonlinear factor $H(Q)=\frac{H_0}{Q_{eq,i}^2(1-\alpha\frac{Q^2}{Q_{eq,i}^2})}$ (based on their Eq.72). For the MAOStress protocol, the stress input waveform is generated as $\sigma_{21}=\sigma_0\cos(\omega t+\phi)$, where $\phi$ is set to zero for the MAOStress protocol. Replacing $H$ and $\sigma_{21}$ in Eq. \ref{eq:3StressTensorSSTNM}, the coupled differential equations in Eq. \nolinebreak\ref{eq:3A-Evolution-set} can be rearranged into a single differential equation
\begin{equation}
    \frac{dA_{11}}{dt}=-\frac{1}{\tau_0}\Biggl[ \frac{(A_{11}-1)+2\frac{\sigma_0^2}{H_0^2}(1-\alpha (A_{11}+2))^2 [\tau_0\omega \sin (\omega t+\phi)\cos (\omega t+\phi)- \cos^2 (\omega t+\phi)]}{1+\frac{2\sigma_0^2}{H_0^2} \alpha(1-\alpha tr\mathbf{A})\cos ^2 (\omega t +\phi)}\Biggr].
    \label{A11Evolution}
\end{equation}
Moreover, the equations can also be rearranged to compute the strain rate $\dot{\gamma}(t)$ as
\begin{equation}
    \dot{\gamma}=\frac{\sigma_0}{\tau_0 H_0}(\alpha(A_{11}+2)-1)[\tau_0\omega\sin(\omega t+\phi)-\cos(\omega t+\phi)]-\frac{\sigma_0}{H_0}\alpha \cos (\omega t+\phi)\frac{dA_{11}}{dt}.
    \label{gammadotEvolution}
\end{equation}
 After solving for the time-dependence of $A_{11}$ using Eq. \nolinebreak\ref{A11Evolution}, $\dot{\gamma}(t)$ can be directly computed from Eq. \nolinebreak\ref{gammadotEvolution}, and $\gamma(t)$ was computed by integrating $\dot{\gamma}(t)$ .
The equilibrium modulus $G_0$ can be calculated as $G_0=\frac{H_0}{1-3\alpha}$. The numerical solution at every stress amplitude and frequency is obtained by solving Eq. \ref{gammadotEvolution} with an initial condition $A_{11}(t=0)=1$ and with $\sigma_{21}=\sigma_0\cos( \omega t+\frac{\pi}{2})$ from $t=0$ to $t=15\frac{2\pi}{\omega}$(15 cycles). We choose $\phi=\frac{\pi}{2}$ to start with an equilibrium state $A_{11}(t=0)=1$. The first 12 cycles are not processed and are used to reach steady-alternance of the system. The last three cycles are processed with a Discrete Fourier Transform to obtain the strain harmonics. Next, similar to the experimental protocol above, a stress amplitude sweep is run at every frequency of interest to extract the MAOStress material functions by fitting the harmonics to the expansion in Eq. \ref{eq:3 harmnonics expansion}. 

\begin{figure}

 \includegraphics{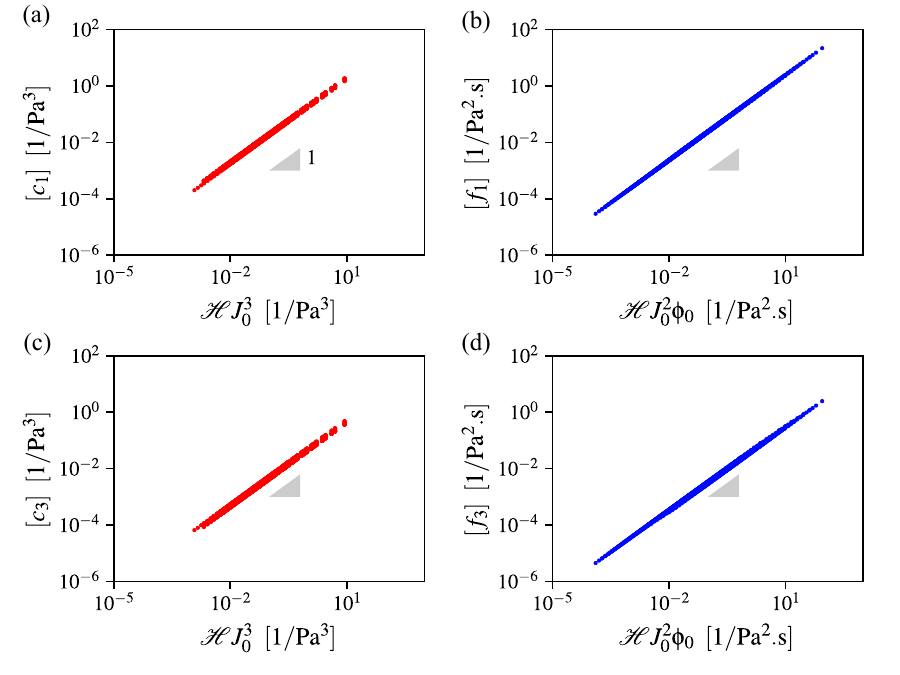}%
 \caption{Scaling of the SSTNM material functions. The scatter of simulation results for different model parameters shows that the numerical simulations follow the analytical solution from Lennon et al. interrelations\cite{Lennon2020MediumExamples} for all the attempted conditions. \label{ScalingSSTNM}}%
 \end{figure}

To test the scaling of the MAOStress material functions with $J_0$, $\tau_0$, a scatter plot was generated by simulating the model with various parameters as shown in Figure \ref{ScalingSSTNM}. The nonlinear coefficient $\mathcal{H}$ was defined by Bharadwaj et al. as $\mathcal{H}=\frac{\partial \ln H(Q)}{\partial \ln Q}|_{Q=Q_{eq}}=\frac{6\alpha}{1-3\alpha}$. The MAOStrain material functions scale linearly with $\mathcal{H}$, and the same was observed for the MAOStress material functions here. Through empirical observation, the material functions $[c_1]$, $[f_1]$, $[c_3]$, and $[f_3]$ were observed to scale  with $J_0^3$, $J_0^2\phi_0$, $J_0^3$, and $J_0^2\phi_0$, respectively. 
 
\subsection*{Analytical Solution Using Lennon et al. Interrelations}
The analytical solutions from the Lennon et al. interrelations corroborate the numerical solution of the SSTNM model. The interrelations (see their Eqs.B3-4) convert MAOStrain material functions to MAOStress material functions, assuming a Volterra series expansion for the stress as a function of strain. The MAOStrain analytical solutions were derived by Bharadwaj et al. in 2017 (see their Eqs.51-54). Using the MAOStrain analytical solution with the Lennon et al. interrelations, the analytical solution of the MAOStress material functions for the SSTNM model was obtained using Wolfram Mathematica symbolic math calculator as

\begin{subequations}\label{eq:3SSTNMSsol}
\begin{align}
 [c_1]&=-\frac{1}{6}\mathcal{H}J_0^3\frac{1+5\mathrm{De}^2}{1+4\mathrm{De}^2}  \\
[f_1]&=-\frac{1}{12}\mathcal{H}J_0^2\phi_0\frac{3+11\mathrm{De}^2}{1+4\mathrm{De}^2} \\
[c_3]&=-\frac{1}{18}\mathcal{H}J_0^3\frac{1+3\mathrm{De}^2}{1+4\mathrm{De}^2} \\
[f_3]&=-\frac{1}{12}\mathcal{H}J_0^2\phi_0\frac{1+5\mathrm{De}^2}{1+4\mathrm{De}^2}.
\end{align}
\end{subequations}

 \begin{figure}
 \includegraphics{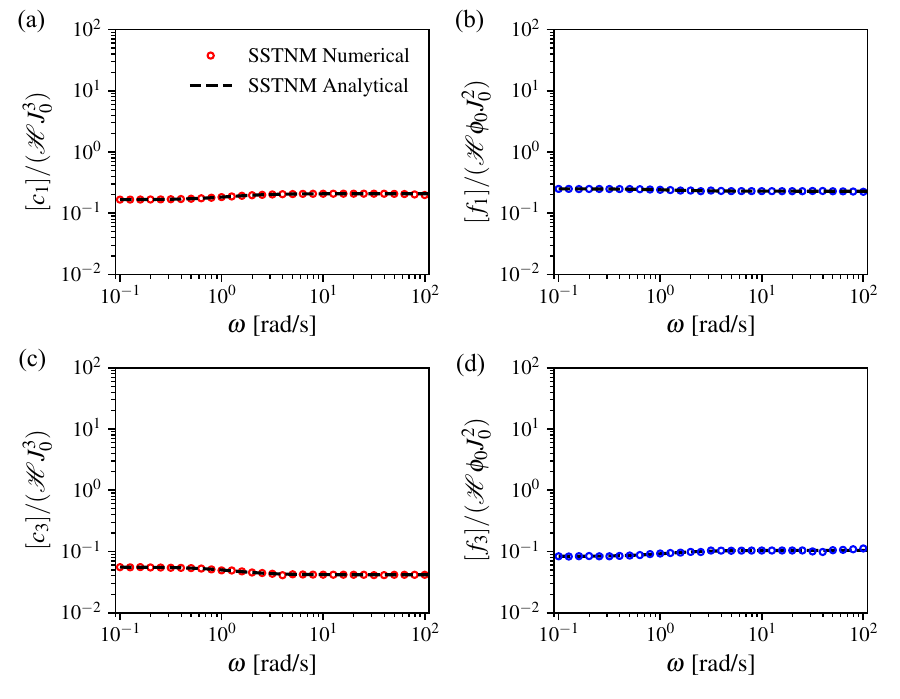}%
 \caption{Intrinsic material functions prediction for SSTNM model. The symbols are for the numerical simulation results, and the dashed lines are obtained analytically from converting the MAOStrain analytical results in \cite{Bharadwaj2017AShear} using the Lennon conversion relations \cite{Lennon2020MediumExamples}. \label{ComparingSSTNMLennon}}%
 \end{figure} 
 Figure \ref{ComparingSSTNMLennon} shows that the numerical and analytical solutions agree exactly, and that the analytical solution predicts the small dependence on the frequency observed. This agreement means that the Volterra series expansion assumed by the Lennon et al. interrelations applies to the SSTNM model. 

To further justify using this scaling in the normalization used for the experimental data, we derived the MAOStress material functions of the MAOS TSS model using the Lennon interrelations. The analytical results of the MAOS TSS models were derived by Martinetti et al. in 2018 (see their Eq. 28). The resulting MAOStress material functions of the MAOS TSS model are:
\begin{subequations}\label{eq:3MAOSTSSsol}
\begin{align}
    [c_1] &=-9AJ_0^3\frac{\mathrm{De}^2}{1+4\mathrm{De}^2} \\
    [f_1] &=-\frac{9}{2}AJ_0^2\phi_0\frac{1+3\mathrm{De}^2}{1+4\mathrm{De}^2} \\
    [c_3] &=-AJ_0^3\frac{\mathrm{De}^2}{1+4\mathrm{De}^2} \\
    [f_3] &=-\frac{3}{2}AJ_0^2\phi_0\frac{1+3\mathrm{De}^2}{1+4\mathrm{De}^2}.
\end{align}
\end{subequations}
The MAOS TSS and the SSTNM models reveal the same scaling with linear material properties. 
\subsection*{Other choices of normalization}
\begin{figure}
 \includegraphics{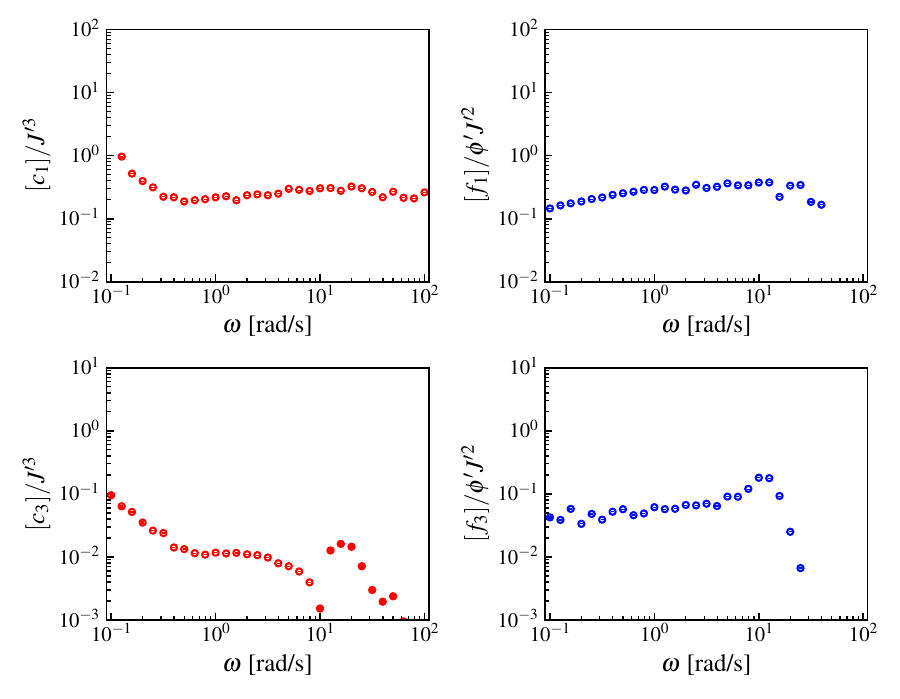}%
 \caption{Alternative normalization of the nonlinear material functions. 
 The normalized intrinsic first harmonic (a) elastic $[c_1]/J^{\prime 3}$and (b) viscous $[f_1]/\phi^{\prime} J^{\prime 2}$ MAOStress material functions show that PVA-Borax stiffens and thickens under weakly-nonlinear stress forcing, with a near constant magnitude ($\approx 0.2$). The normalized intrinsic third harmonic (a) elastic $[c_3]/J^{\prime 3}$and (b) viscous $[f_3]/\phi^{\prime} J^{\prime 2}$ are similar to the ones reported in Figure \ref{fig:3MAOSsweep}. Open symbols represent negative values, and closed symbols represent positive values.}%
 \label{fig:3MAOSsweepnorm2}
 \end{figure}

In the results section, we discussed the reasoning for why these models are predicting this scaling. Moreover, we attempted to generalize the normalization for materials and models that do not have a single relaxation mode by replacing $J_0$ and $\phi_0$ by $J^{\prime}$ and $\phi^{\prime}$, respectively. Figure \ref{fig:3MAOSsweepnorm2} shows that this normalization does reduce the complexity of the MAOStress material function representation for the PVA-Borax sample. 

\section*{Appendix C: Measurement of MAOStrain material functions}
\begin{figure}[h!]
 \includegraphics{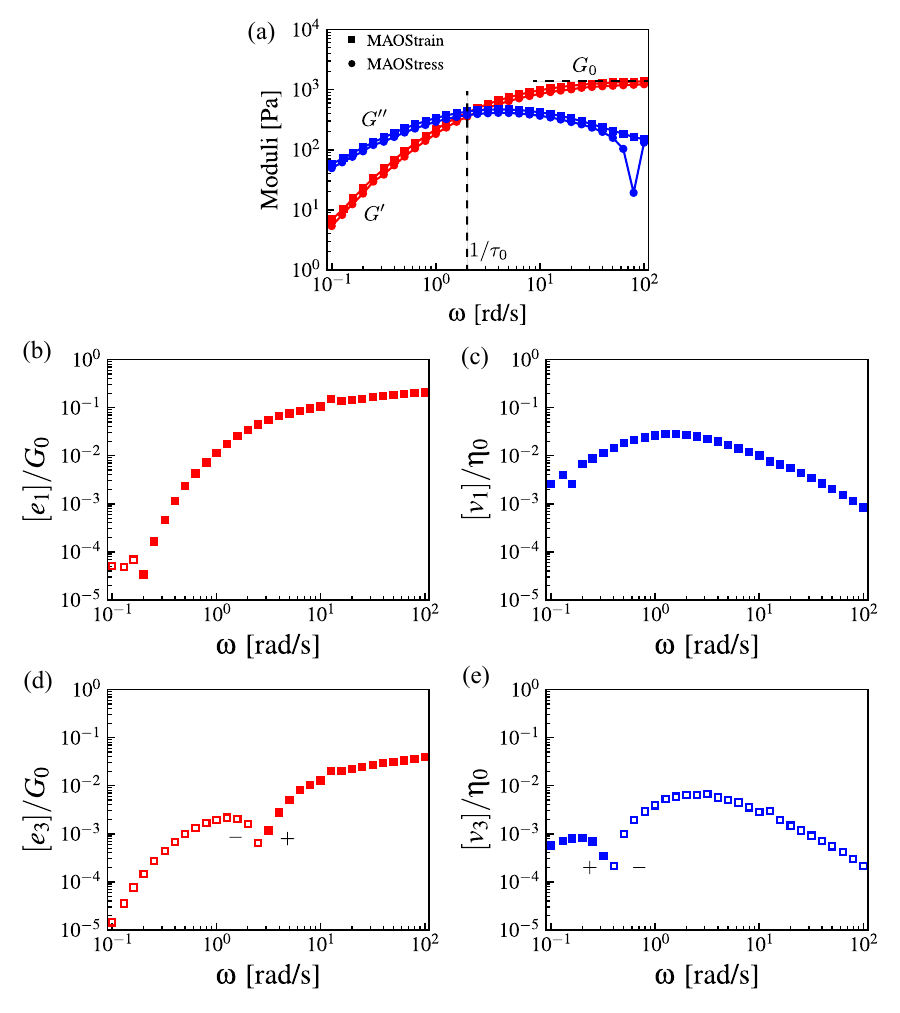}%
 \caption{MAOStrain material functions. (a) LVE moduli $G^{\prime}$ and $G^{\prime\prime}$ for PVA-Borax from the MAOStrain experiments on the ARES-G2 (Squares) and the MAOStress experiments on the MCR 702 (Circles). The cross-over time scale and plateau modulus are plotted based on the lognormal spectrum fit. The normalized intrinsic MAOStrain material functions (b) $[e_1]$, (c) $[v_1]$, (d) $[e_3]/G_0$, and (e) $[c_3]/\eta_0$ are plotted as a function of frequency. Open symbols for negative values and closed symbols for positive values. }
 \label{fig:3MAOStrain}%
 \end{figure}
   \begin{figure}[h!]
 \includegraphics{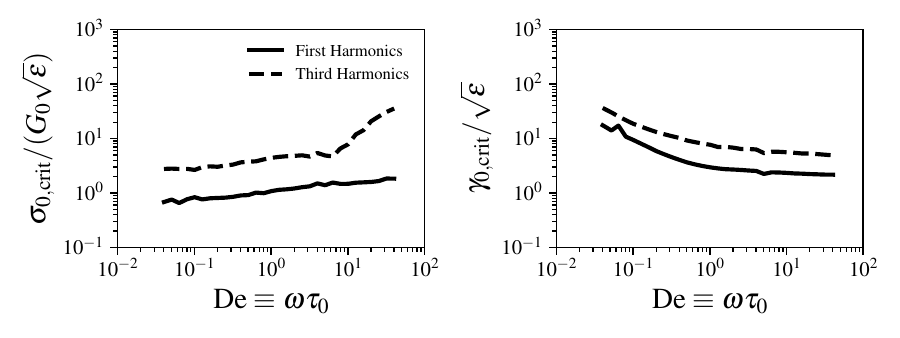}%
 \caption{The (a) critical stress and (b) critical strain based on the magnitude of the first and third harmonic deviations.}
 \label{fig:3MAOStrainFirstAndThird}%
 \end{figure}

The MAOStrain material functions of PVA-Borax were measured on the strain-controlled ARES-G2 rheometer. These material functions were measured in previous work, but we repeat them here for consistency and to be able to directly compare them to the MAOStress results. The material was prepared as outlined in the Materials and Methods section above. Oscillatory strain amplitude sweeps ($\gamma=\gamma_0\sin \omega t$) were conducted using a parallel plate geometry (d = 25mm, 0.5mm gap) at $25^\circ \mathrm{C}$ maintained by a Peltier system in the lower plate. A parallel plate geometry allows for faster material loading at the expense of a spatially inhomogeneous strain field. We use the single-point parallel-plate corrections for MAOStrain derived by Bharadwaj et al. (exact corrections) to account for the inhomogeneity\cite{Bharadwaj2015Single-pointShear}. The sample is loaded on the instrument by gravity pouring it over the bottom plate. Then, the upper geometry is slowly lowered onto the sample with a maximum normal force of 1 N to prevent stress build-up in the sample. Once the upper geometry was reduced enough to overfill the gap slightly, the sample was trimmed to ensure a proper fill. Then, silicone oil was applied on the outer edge to prevent evaporation loss. Before starting any tests, the sample was allowed to relax for 10 minutes to remove memory from deformations during the loading procedure.

The full frequency range ($\omega=[0.1-100] \mathrm{rad/s}$)  was tested in two loadings. We used $\gamma_0=[0.1,120]\%$ for $\omega=[0.1-1] \mathrm{rad/s}$, and $\gamma_0=[0.1,100]\%$ for $\omega=[1.0-100.0]\mathrm{rad/s}$. Moreover, we used 50 points per decade of $\gamma_0$ to get good resolution and to ensure that the steps in strain amplitude are small, such that the equilibration time from one amplitude to the other is low. A Discrete Fourier Transform is used to calculate the stress harmonics of the stress oscillatory waveform $\sigma(t)$. The harmonics are defined using this expansion: 
\begin{equation}
\label{eq:3 Fourier expansionStrain}
    \sigma(t;\gamma_0,\omega)=\sum_{n,\mathrm{odd}} \sigma_n^{\prime}(\gamma_0,\omega)\sin n\omega t+\sigma_n^{\prime\prime}(\gamma_0,\omega)\cos n\omega t.
\end{equation}
Similar to previous work, the first and third stress harmonics can be expanded as: 
\begin{subequations}
\label{eq:3 harmnonics expansionStrain}
\begin{eqnarray}
\sigma_1^{\prime}(\gamma_0,\omega)=G^{\prime}\gamma_0 + [e_1]\gamma_0^3+ \mathcal{O}(\gamma_0^5)\\
\sigma_1^{\prime\prime}(\gamma_0,\omega)=G^{\prime\prime}\gamma_0 + [v_1]\gamma_0^3 \mathcal{O}(\gamma_0^5)\\
\sigma_3^{\prime}(\gamma_0,\omega)= [e_3]\gamma_0^3+ \mathcal{O}(\gamma_0^5)\\
\sigma_3^{\prime\prime}(\gamma_0,\omega)= [v_3]\omega\gamma_0^3+ \mathcal{O}(\gamma_0^5).
\end{eqnarray}
\end{subequations}
The MAOStrain material functions $G^{\prime}$, $G^{\prime \prime}$, $[e_1]$, $[v_1]$, $[e_3]$, and $[v_3]$ were extracted from the amplitude sweeps with the same fitting procedure used for the MAOStress material functions described in section II and III. 

Figure \ref{fig:3MAOStrain} shows the MAOStrain material functions as a function of frequency. Moreover, Figure \ref{fig:3MAOStrain} compares the linear material functions from MAOStrain and MAOStress, showing that the two procedures resulted in identical linear material functions. The MAOStress moduli were obtained by converting the measured compliances, using the interchangeability relationships Eqs.\ref{eq:3LinInter1} and \ref{eq:3LinInter2}. The MAOStrain material functions were used to compute the critical strain for the onset of nonlinearity using Eqs. \ref{eq:3critStrainFirstHarmonics} and \ref{eq:3critStrainThirdHarmonics}. Moreover,  Figure \ref{fig:3MAOStrainFirstAndThird} compares the critical stress and strain based on the first and third harmonics.

\bibliography{mendeley,references}

\end{document}